\documentclass[a4paper,11pt,usenames,dvipsnames,table]{article}

\usepackage{jheppub} 
\usepackage[T1]{fontenc} 
\usepackage{float}
\usepackage[usenames,dvipsnames]{xcolor}
\usepackage{mathtools}
\usepackage[]{mdframed}
\usepackage{amssymb}
\usepackage{amsthm} 
\usepackage{xfrac}

\usepackage{booktabs} 

\title{Renormalization Group is the Principle \\behind the Holographic Entropy Cone}

\author{Bart{\l}omiej Czech and Sirui Shuai}

\affiliation{Institute for Advanced Study, Tsinghua University, Beijing 100084, China}

\emailAdd{bartlomiej.czech@gmail.com}
\emailAdd{siruishuai@gmail.com}

\abstract{We show that every holographic entropy inequality can be recast in the form: `some entanglement wedges reach deeper in the bulk than some other entanglement wedges.' When the inequality is saturated, the two sets of wedges reach equally deep. Because bulk depth geometrizes CFT scales, the inequalities enforce and protect the holographic Renormalization Group.}

\begin{document} 
\maketitle
\flushbottom

\section{Introduction}
\label{sec:introduction}

A major theme in the AdS/CFT correspondence \cite{Maldacena:1997re,Witten:1998qj} is that the gravitational side of the duality is an information theoretic rendering of the field theory side, geometrically organized \cite{VanRaamsdonk:2010pw}. Key tenets of this information theoretic rendering are:
\begin{enumerate}
\item The Ryu-Takayanagi proposal \cite{Ryu:2006bv, rt2}: von Neumann entropies of spatial subregions are manifested as areas of extremal surfaces plus $\mathcal{O}(1/N)$ corrections. We refer to the requisite surfaces as {\bf RT surfaces} after Ryu and Takayanagi.
\item Subregion duality \cite{Czech:2012bh,Dong:2016eik}: the physics in a boundary spatial subsystem $A$ solely controls the physics in a bulk subsystem called the entanglement wedge ${\rm EW}(A)$, which is demarcated by the Ryu-Takayanagi extremal surface.
\item Low energy bulk physics as an erasure correcting code \cite{Almheiri:2014lwa}: when entanglement wedges of different subsystems overlap, different parties can share control over the same \emph{code subspace}. That is, overlapping ${\rm EW}(X_i)$ indicate that parties in boundary regions $X_i$ share access to the underlying bulk physics, which in turn is protected against erasures of $\overline{X_i}$ (the complement of $X_i$).
\item Holographic Renormalization Group (RG) \cite{Balasubramanian:1999jd}: the bulk geometry is organized into layers that correspond to CFT scales, with the near-boundary regions encoding the CFT ultraviolet (UV) physics and the central region encoding the infrared (IR). 
\end{enumerate}
In this list, points 1.-3. are directly controlled by quantum entanglement in the CFT---including the erasure correcting property, whose geometric contours are decided by the RT surfaces. Point 4. may seem more distant, but in fact the Renormalization Group is closely interlaced with quantum entanglement. This is evident in the entanglement-based proofs of the $c$- and $F$- and $a$-theorems \cite{Casini:2004bw,Myers:2010xs,Casini:2012ei,Casini:2016udt,Casini:2017vbe}, measures of UV-IR entanglement \cite{Balasubramanian:2011wt} and---perhaps most visually striking---in the Multi-scale Entanglement Renormalization Ansatz \cite{Vidal:2006sxo,Vidal:2008zz}.

One statement, which pins together quantum erasure correction, holographic RG and geometric properties of RT surfaces is {\bf entanglement wedge nesting}:
\begin{align}
X_i \subset X_j \quad 
& \Longrightarrow \quad {\rm EW}(X_i) \subset {\rm EW}(X_j)
\label{ewn1} \\
X_i \cap X_j = \emptyset \quad 
& \Longrightarrow \quad {\rm EW}(X_i) \cap {\rm EW}(X_j) = \emptyset
\label{ewn2}
\end{align}
Assertions~(\ref{ewn1}-\ref{ewn2}) are equivalent under the exchange $X_j \leftrightarrow \overline{X_j}$. Both are necessary conditions for holographic erasure correction to work because---focusing on (\ref{ewn1})---data protected against the erasure of $\overline{X_i}$ is automatically also protected against the erasure of $\overline{X_j} \subset \overline{X_i}$. From an RG point of view, (\ref{ewn1}) says that ${\rm EW}(X_j)$ can reach deeper in the bulk than ${\rm EW}(X_i)$ because $X_j$ encompasses CFT scales that are more infrared than those encapsulated in $X_i$. As a geometric statement about RT surfaces, wedge nesting has a very crisp meaning: it demands that Ryu-Takayanagi surfaces for $X_i$ and $X_j$ cannot cross.

While entanglement nesting naturally combines the perspectives of quantum entanglement, renormalization group and geometry, the manner in which it does so is somewhat coarse. Because enlarging $X_j$ to obtain $X_i \supset X_j$ entails adding UV degrees of freedom, ${\rm EW}(X_i)$ is guaranteed to be larger than ${\rm EW}(X_j)$ in its boundary extent but may or may not reach deeper in the bulk. It would be desirable to formulate analogues or extensions of entanglement wedge nesting, which apply to regions of \emph{equal} boundary support (independent of UV entanglement) and specifically probe the bulk depth of entanglement wedges (depend on exclusively IR entanglement). Ideally, such statements would also tell us more detailed information about \emph{access structures} of holographic erasure correcting codes---which combinations of parties can / cannot decode the same logical information. In this paper, we explain that such statements are already known and well studied, albeit in a somewhat disguised form. They comprise {\bf the holographic entropy cone} \cite{Bao:2015bfa}. 

The holographic entropy cone collects linear inequalities, which are obeyed by areas of RT surfaces by virtue of their minimality.\footnote{This assumes time reversal symmetry. In this paper we sidestep the issue of whether the same inequalities hold when the bulk is not time reflection-symmetric and entanglement entropies are computed by the HRT proposal \cite{hrt, maximin}. Progress on this question includes \cite{Czech:2019lps, Grado-White:2024gtx, Grado-White:2025jci, Grimaldi:2025jad}.} The cone terminology reflects the fact that each inequality is saturated on a hyperplane in entropy space (the vector space of hypothetical values of subsystem entropies) and, therefore, the set of holographically admissible entropy assignments forms a cone. The inequalities that bound the holographic entropy cone are famously intricate and difficult to parse \cite{HernandezCuenca:2019wgh, Hernandez-Cuenca:2023iqh, Czech:2023xed}; see ineq.~(\ref{rp2ineqs}) for an example. It has been a long standing challenge to interpret them in physical or operational terms; see \cite{Czech:2021rxe, Czech:2025jnw} for earlier partial results. We venture to propose an interpretation.

We argue that the holographic entropy inequalities---the bounding facets of the holographic entropy cone---enforce and protect holographic RG. In essence, they stipulate that the more infrared regions' entanglement wedges reach deeper into the bulk than do entanglement wedges of ultraviolet regions, with the saturation of an inequality implying precisely equal depth. When this rule of thumb is further refined using technical tools, one discovers that saturating an inequality eliminates many otherwise reasonable access structures of the quantum erasure correcting code \cite{Czech:2025jnw}. Thus, the non-negative quantity defined by each inequality is an order parameter for a certain type of erasure correction.

\paragraph{Setup}
We consider time reflection-symmetric, asymptotically AdS geometries, which are dual to states of a conformal field theory (CFT). At leading order in Newton's constant $G_N$, the entanglement entropy of a boundary region $X$ is given by \cite{Ryu:2006bv, rt2}:
\begin{equation}
S_X = (4 G_N)^{-1} \min_{\Sigma \cup X = \partial E(X)} {\rm Area}(\Sigma)
\label{eq:rt}
\end{equation}
The domain of minimization spans those surfaces $\Sigma$, which together with $X$ itself bound a submanifold $E(X)$ of the time reversal-symmetric bulk slice; this is known as the homology condition. The submanifold $E(X)$ stipulated in (\ref{eq:rt}) is the main protagonist of this paper. We admonish the reader to distinguish $E(X)$ from the entanglement wedge ${\rm EW}(X)$: ${\rm EW}(X)$ is the bulk domain of dependence of $E(X)$ and $E(X)$ is the restriction of ${\rm EW}(X)$ to the spatial slice fixed by time reversal symmetry.

Proposal~(\ref{eq:rt}) implies that the CFT entanglement entropies satisfy infinitely many inequalities \cite{Bao:2015bfa} of the schematic form
\begin{equation}
{\rm LHS} \equiv \sum_i \alpha_i S({X_i}) \geq \sum_j \beta_j S({Y_j}) \equiv {\rm RHS},
\label{eq:schematic}
\end{equation}
where all coefficients $\alpha_i$ and $\beta_j$ are positive. With the exception of subadditivity $S_A + S_B \geq S_{AB}$, every primitive (maximally tight) holographic inequality admits quantum mechanical states, which violate it. Thus, inequalities~(\ref{eq:schematic}) are strictly consequences of the RT proposal and represent necessary conditions for its validity.

\subsection{Overview of results}
We argue that the inequalities serve the following purpose: to enforce and protect the layered organization of the bulk AdS geometry by CFT scale, with infrared (IR) physics residing deeper in the bulk. This is encapsulated by the following statements, which apply to \emph{every} inequality~(\ref{eq:schematic}):\footnote{The inequality has to be written in a \emph{balanced} form \cite{He:2020xuo}. We explain this point further in Section~\ref{sec:inclusion}.}
\begin{align}
\textrm{Inclusion:} &\quad \textrm{For every $X_i$ there is a $Y_j$ that contains it.} 
\label{eq:lemma} \\
\textrm{Collective Inclusion:} &\quad \!\cup_i\! E(X_i) \subseteq \cup_j E(Y_j) 
\label{eq:inclusion} \\
\!\!\!\textrm{Saturation:} &\quad \textrm{if~~${\rm LHS} = {\rm RHS}$~~then~~} \cup_i\! E(X_i) = \cup_j E(Y_j)
\label{eq:saturation}
\end{align}
In the language of the Renormalization Group, claim~(\ref{eq:lemma}) states that regions on the RHS of inequality~(\ref{eq:schematic}) contain infrared data, which are not present on the LHS. Claim~(\ref{eq:inclusion}) follows directly from (\ref{eq:lemma}) and wedge nesting~(\ref{ewn1}). However, because all holographic entropy inequalities have $\cup_i X_i = \cup_j Y_j$, (\ref{eq:inclusion}) is strictly a statement about \emph{bulk depth}. It says that the entanglement wedges of the more infrared RHS regions collectively reach deeper in the bulk than do entanglement wedges of the relatively ultraviolet LHS regions. Finally, claim~(\ref{eq:saturation}) says that additional depth is reached only if the inequality is not saturated. In this sense, the inequality functions as an order parameter for the infrared object $\cup_j E(Y_j)$ reaching additional bulk depth beyond $\cup_i E(X_i)$. One illustration of these heuristics is Figure~\ref{fig:dihedral_core} in Section~\ref{sec:examples}.

For a preview of how these results work, let us apply them to the subadditivity of entanglement entropy:
\begin{equation}
S(A) + S(B) \geq S(AB)
\label{eq:sa}
\end{equation}
Claims~(\ref{eq:lemma}-\ref{eq:inclusion}) are manifest. Applying~(\ref{eq:saturation}), we find that if $I(A:B) = S(A) + S(B) - S(AB) = 0$ then $E(AB) = E(A) \cup E(B)$. Thus, the mutual information $I(A:B)$ is the order parameter for the infrared object $E(AB)$ to reach deeper into the bulk than $E(A)$ and $E(B)$ do. We illustrate these findings in Figure~\ref{fig:sat_sa} in Section~\ref{sec:sa}.

Interesting applications of result~(\ref{eq:saturation}) concern the strong subadditivity as well as the monogamy of mutual information. In both cases, we obtain strong statements about the structures of the underlying entanglement wedges; viz. equations~(\ref{eq:ssaresult}) and (\ref{mmiresult1}). 

Results~(\ref{eq:lemma}-\ref{eq:saturation}) are derived in Section~\ref{sec:rg}.

\paragraph{A further refinement}
In prior work \cite{Czech:2025jnw}, Wang and we showed that the saturation of an inequality ${\rm LHS} = {\rm RHS}$ is a no-go condition for certain access structures in a holographic erasure correcting code. Applied directly to inequalities~(\ref{eq:schematic}), those results are distinct and logically independent from claim~(\ref{eq:saturation}). However, applying \cite{Czech:2025jnw} to the tautologically extended inequality ${\rm LHS} + S(Y_j) \geq {\rm RHS} + S(Y_j)$ reveals a wealth of further constraints on holographic erasure correction, which apply when (\ref{eq:schematic}) is saturated. These constraints provide an improvement on (\ref{eq:saturation}), which in many cases functions at the level of individual $E(Y_j)$ and not their union:
\begin{equation}
\textrm{if\,~${\rm LHS} = {\rm RHS}$~~then~~}E(Y_j) = \cup_{\textrm{few terms}}\, E(X_i)
\label{eq:refined}
\end{equation}
Technical versions of (\ref{eq:refined}) are given in (\ref{cyclicgen}), (\ref{eq:toricrefined}) and (\ref{rpref1}-\ref{rpref3}). These refinements of result~(\ref{eq:saturation}) are the subject of Section~\ref{sec:main}.

\section{Inequalities and holographic RG}
\label{sec:rg}

Here we derive claims~(\ref{eq:lemma}) and (\ref{eq:saturation}). Claim~(\ref{eq:inclusion}) does not require an independent justification because it follows directly from (\ref{eq:lemma}) and entanglement wedge nesting~(\ref{ewn1}). 

\subsection{Inclusion}
\label{sec:inclusion}
Statement~(\ref{eq:lemma}) can be justified in two ways. The simpler way, which we present here, relies on a conjectured property of all inequalities~(\ref{eq:schematic}), which was recently proposed in \cite{Grimaldi:2025jad}. Another argument exploits an auxiliary object called a contraction map, which is used in the conventional technique of proving holographic inequalities \cite{Bao:2015bfa}. We explain contraction proofs in Appendix~\ref{sec:contraction} and give a contraction-based proof of (\ref{eq:lemma}) in Appendix~\ref{sec:anotherproof}. 

It is conceivable that some hitherto unknown entropy inequality might be holographically valid \emph{and} disprove the conjecture of \cite{Grimaldi:2025jad} \emph{and} have no proof by the contraction method, though Reference~\cite{Bao:2025sjn} argued that no such inequality exists. We return to this theoretical possibility in the Discussion.

\paragraph{Majorization test} The conjecture in \cite{Grimaldi:2025jad} stipulates that every holographic inequality passes a {test,} which we presently explain. We use the monogamy of mutual information \cite{Hayden:2011ag}
\begin{equation}
S(AB) + S(BC) + S(AC) \geq S(A) + S(B) + S(C) + S(ABC)
\label{eq:mmi}
\end{equation}
to illustrate the conjecture. 

The test presumes that the inequality is written in a \emph{balanced} form---one where every region appears the same number of times on the LHS and on the RHS.\footnote{Working with global pure states and assuming Haag duality holds (see e.g. \cite{Casini:2019kex} for scenarios where it does not), it is possible to rewrite a balanced inequality in an unbalanced form by exploiting $S(X) = S(\overline{X})$. Such rewritings can sometimes be useful \cite{Czech:2023xed}, but not for the purposes of this paper.} Example~(\ref{eq:mmi}) manifestly satisfies the balanced requirement. Reference~\cite{He:2020xuo} showed that \emph{every} holographic entropy inequality can be written in a balanced form.

For every indivisible region such as $A$, obtain a \emph{reduced inequality} by dropping all terms that do not contain the said region. In the monogamy example, the $A$-reduced inequality is $S_{AB} + S_{AC} \geq S_{ABC} + S_A$. Based on the reduced inequality, form two vectors---$x$ from its left hand side and $y$ from the right. The components of the vectors correspond to terms of the reduced inequality, and are obtained by replacing each indivisible region with a corresponding variable (here denoted in lowercase) in a sum. For the $A$-reduced inequality, this gives $x = (a+b, a+c)$ and $y = (a+b+c, a)$. Then \emph{for every choice of} positive values of $a,b,c\ldots$, vectors $x$ and $y$ satisfy the majorization condition $x \prec y$. In particular, labeling the components of $x$ and $y$ from largest to smallest as $x_1, x_2 \ldots$ and $y_1, y_2 \ldots$, we have:
\begin{equation}
\forall\,k: \quad \sum_{i=1}^k x_i \leq \sum_{j=1}^k y_j
\label{eq:majorization}
\end{equation}
In the example, assuming without loss of generality that $b > c$, we have $(x_1, x_2) = (a+b, a+c)$ and $(y_1, y_2) = (a+b+c, a)$ and the claim is manifestly true.

It suffices to prove claim~(\ref{eq:lemma}) for those $X_i$, which are not contained in any other $X_{i'}$. To do so, apply the test using any constituent of $X_i$ in the step where terms are dropped. (In other words, reduce the inequality on any constituent of $X_i$.) In the step where variables are introduced, set the variables that correspond to the constituents of $X_i$ to be parametrically larger than all others. Since $X_i$ is not contained in any other $X_{i'}$, this guarantees that the term which descends from $X_i$ is the largest component of $x$. Then~(\ref{eq:majorization}) with $k=1$ implies that the largest component of $y$ must contain all the parametrically large variables. Therefore, its parent RHS term---call it $Y_{j(i)}$---must contain all the constituents of $X_i$.

Returning to the example, to deduce that some RHS term in (\ref{eq:mmi}) contains $AB$, we would choose $a,b \gg 1$ and $c \ll 1$ and identify $ABC \supset AB$ as the term from which $y_1 \geq a+b$ descends.
\medskip

\paragraph{Superbalance and different notions of depth} In the preceding argument, only a subset of RHS terms appear relevant. In particular, in the monogamy example, claim~(\ref{eq:inclusion}) is validated by term $S(ABC)$ alone. The other RHS regions, $A$ and $B$ and $C$, are ostensibly irrelevant.

One retort is that they are necessary to render the difference $({\rm LHS} - {\rm RHS})$ UV-finite in the CFT. Indeed, UV-finiteness---known under the monicker `superbalance'---is a proven property of all holographic entropy inequalities \cite{He:2020xuo}. But superbalance entails a further corollary: the inequality must pass the majorization test of \cite{Grimaldi:2025jad} even if we rewrite it using a different choice of purifier. We explain this presently.

In the monogamy example, introduce a purification $O$ such that the state on $ABCO$ is pure. Assuming complementary regions have equal entropies, we may then rewrite monogamy as:
\begin{equation}
S(AB) + S(AO) + S(BO) \geq S(A) + S(B) + S(ABO) + S(O)
\label{eq:mmi2}
\end{equation}
This rewriting treats $C$ as the purifier---that is, we exploit $S(X) = S(\overline{X})$ to rewrite every term in the way that excludes $C$. When we change purifiers in a superbalanced inequality, the rewritten inequality still passes the test of \cite{Grimaldi:2025jad} and, consequently, respects claims~(\ref{eq:lemma}) and (\ref{eq:inclusion}). This is not true for inequalities, which are not superbalanced.

After changing the purifier, different terms validate (\ref{eq:lemma}). In inequality~(\ref{eq:mmi2}), it is region $ABO$ that makes the lemma true---but its origin in (\ref{eq:mmi}) is $C=\overline{ABO}$, which appeared irrelevant when the lemma was applied to monogamy in its original form. 

Generally, the notion `deeper in the bulk'---as well as statements~(\ref{eq:lemma}) and (\ref{eq:inclusion})---must be understood with respect to a particular purifier: {\bf `deeper' means closer to the purifier.} If the purifier is connected and the CFT lives on a connected manifold, we may think of the choice of purifier as selecting a point at infinity; the CFT infrared is then \emph{defined} as approaching that point. If the purifier is connected but the CFT lives on a disconnected manifold, we conceptualize bulk depth as approaching the connected component with the purifier. Cases where the purifier is disconnected defy such pithy verbal descriptions. However, statement~(\ref{eq:inclusion}) always implicitly defines a notion of depth: the bulk region $\big(\cup_j\! E(Y_j)\big) \setminus \big(\cup_i\! E(X_i)\big)$ lies deeper than $\cup_i E(X_i)$ does.

\subsection{Saturation}
\label{sec:saturationclaim}
We work with inequalities~(\ref{eq:schematic}) that can be proven by the contraction method \cite{Bao:2015bfa}; if a valid inequality not provable by contraction exists, our results do not apply to it. We review proofs by contraction in Appendix~\ref{sec:contraction}, but the argument can be stated without reference to the mechanics of such proofs. The following text is self-contained in that a working knowledge of contraction maps is not necessary.

The essence of proofs by contraction is to construct candidate Ryu-Takayanagi surfaces $\Sigma^{\rm can}_j$ by cutting and gluing segments of actual Ryu-Takayanagi surfaces of regions $X_i$. The candidate surfaces are part of the minimization domain in (\ref{eq:rt}) and constructed so that ${\rm Area}(\Sigma^{\rm can}_j) \equiv 4G_N S^{\rm can}(Y_j)$ satisfy:
\begin{equation}
{\rm LHS} \equiv \sum_i \alpha_i S(X_i) \geq \sum_j \beta_j S^{\rm can}(Y_j) \equiv {\rm RHS}^{\rm can}
\label{eq:schemcand}
\end{equation}
Any such construction proves (\ref{eq:schematic}) because $S^{\rm can}(Y_j) \geq S(Y_j)$ by minimality.

Using the candidate surfaces $\Sigma^{\rm can}_j$ in the definition of the entanglement wedge, we may define candidate homology regions $E^{\rm can}(Y_j)$, which are bounded by $\Sigma^{\rm can}_j \cup Y_j$. Because proofs by contraction obtain $\Sigma^{\rm can}_j$ from the physical Ryu-Takayanagi surfaces $\Sigma_i$ of regions $X_i$, it is obvious that:
\begin{equation}
\cup_j E^{\rm can}(Y_j) \subseteq \cup_i E(X_i)
\label{eq:reverse}
\end{equation}
To confirm~(\ref{eq:saturation}), observe that if ${\rm LHS} = {\rm RHS}$ then (\ref{eq:schemcand}) and $S^{\rm can}(Y_j) \geq S(Y_j)$ must be saturated (hold with equality) for all $j$. This means that $\Sigma^{\rm can}_j$ is the minimizing surface $\Sigma_j$ and $E^{\rm can}(Y_j) = E(Y_j)$. Then combining (\ref{eq:reverse}) with result~(\ref{eq:inclusion}) establishes~(\ref{eq:saturation}). 

The only caveat in this argument is if two surfaces $\Sigma^{\rm can}_j$ and $\Sigma'_j$ simultaneously achieve the minimum in (\ref{eq:rt}). We would then be at a phase transition---a non-generic scenario, in which proposal~(\ref{eq:rt}) is subject to enhanced corrections \cite{Dong:2020iod,Akers:2020pmf}. A common prescription for defining entanglement wedges at a phase transition is to select the smallest possible wedge $E(Y_j) \subseteq E^{\rm can}(Y_j)$, which again confirms~(\ref{eq:saturation}).

We remind the reader that when ${\rm LHS} = {\rm RHS}$, statement~(\ref{eq:saturation}) applies with respect to all choices of purifier. 

\subsection{Examples}
\label{sec:examples}

\paragraph{Strong subadditivity} Inequality
\begin{equation}
S(AB) + S(BC) \geq S(ABC) + S(B)
\label{eq:ssa}
\end{equation}
clearly satisfies claim~(\ref{eq:lemma}). When (\ref{eq:ssa}) is saturated, according to (\ref{eq:saturation}) we have:
\begin{equation}
E(ABC) = E(AB) \cup E(BC)
\label{eq:ssaresult}
\end{equation}

Strong subadditivity is not a primitive (maximally tight) holographic entropy inequality because it is implied by the monogamy of mutual information~(\ref{eq:mmi}) and subadditivity. Indeed, assuming monogamy we have:
\begin{equation}
S(AB) + S(BC) - S(ABC) - S(B) \geq S(A) + S(C) - S(AC) \geq 0
\end{equation}
Thus, strong subadditivity evades the result of \cite{Czech:2025jnw} and it is not superbalanced. Consequently, we do not get to apply (\ref{eq:saturation}) for every choice of purifier in (\ref{eq:ssa}). However, strong subadditivity \emph{is} symmetric under the exchange $B \leftrightarrow O$ (where the state on $ABCO$ is pure) so we have:
\begin{equation}
E(ACO) = E(AO) \cup E(CO)
\label{eq:ssaresult2}
\end{equation}

Conclusion~(\ref{eq:ssaresult})---or rather its contrapositive---was previously established in \cite{May:2021raz}. That analysis was more robust because it did not assume time reversal symmetry and applied to full entanglement wedges, not merely their restrictions to a specific bulk slice. 

\paragraph{Monogamy of mutual information}
We have already seen that monogamy confirms claim~(\ref{eq:lemma}), regardless of the choice of purifier. Applying result~(\ref{eq:saturation}) to monogamy in the form~(\ref{eq:mmi}), we find that its saturation implies:
\begin{equation}
E(ABC) = E(AB) \cup E(BC) \cup E(AC)
\label{mmiresult1}
\end{equation}
When we change purifier and consider monogamy in the form~(\ref{eq:mmi2}), we find that its saturation implies:
\begin{equation}
E(ABO) = E(AB) \cup E(AO) \cup E(BO)
\label{mmiresult2}
\end{equation}
Two other consequences of the saturation of the monogamy of mutual information can be obtained by casting $A$ and $B$ as purifiers. They are identical in form, up to a permutation on region labels $\{A, B, C, O\}$. 

\begin{figure}[t]
    \centering
    \includegraphics[width=0.61\linewidth]{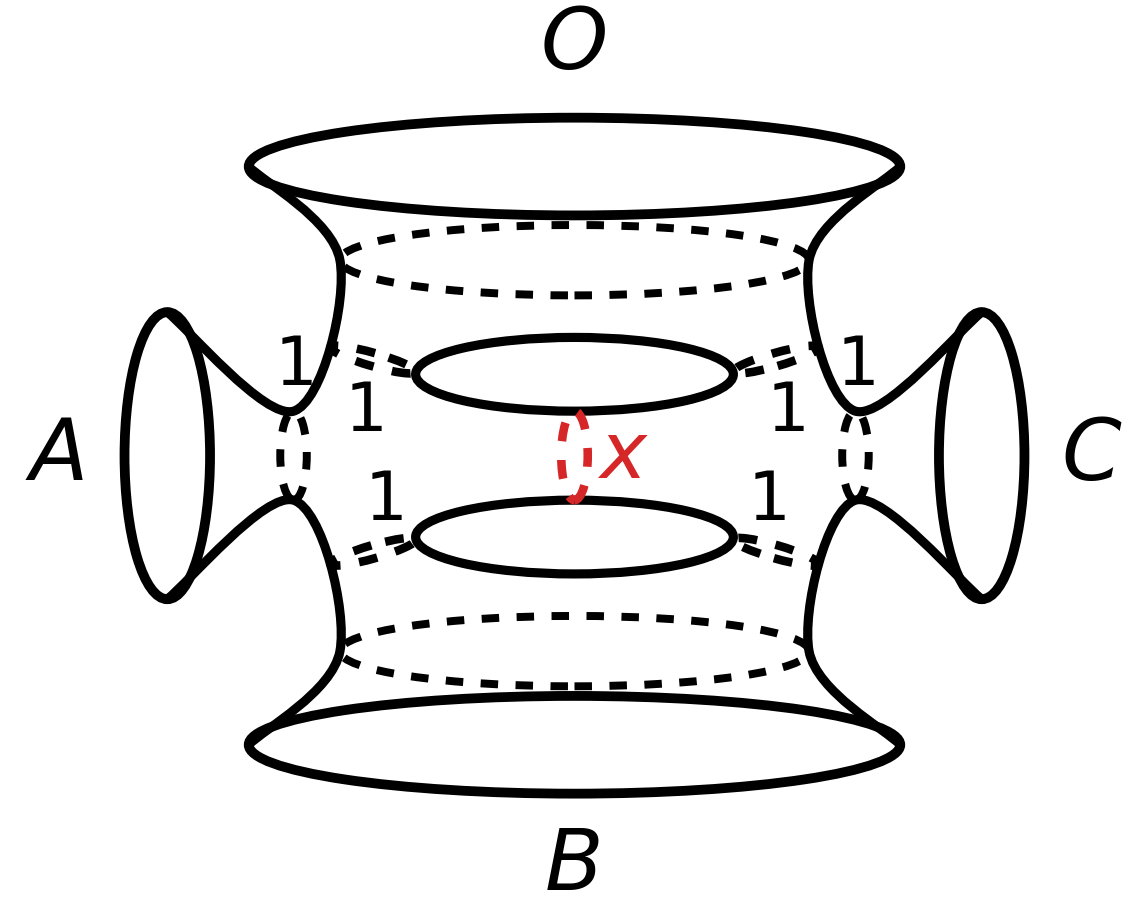}
    \caption{A spatial geometry, in which equations~(\ref{mmiresult1}) and (\ref{mmiresult2}) and their analogues for $E(ACO)$ and $E(BCO)$ hold yet the monogamy of mutual information is not saturated. The example assumes that $x < 1$ and the two large cuts are much larger than 1.}
    \label{fig:counterex}
\end{figure}

\paragraph{Equal depth is not sufficient for saturation} 
According to (\ref{eq:saturation}), `reaching equally deep' with respect to \emph{every purifier} is a necessary condition for the saturation of the inequality. But, as we illustrate in Figure~\ref{fig:counterex}, it is not sufficient. The spatial slice of a holographic geometry shown there does not saturate the monogamy of mutual information~(\ref{eq:mmi}). However, it does respect equations~(\ref{mmiresult1}) and (\ref{mmiresult2}) and their analogues for $E(ACO)$ and $E(BCO)$.

\paragraph{The five-party dihedral inequality} reads \cite{Bao:2015bfa}:
\begin{align}
& S(ABC) + S(BCD) + S(CDE) + S(DEA) + S(EAB) \label{eq:5party} \\
\geq\,\,  & S(AB) + S(BC) + S(CD) + S(DE) + S(EA) + S(ABCDE) \nonumber
\end{align}
Claims~(\ref{eq:lemma}) and (\ref{eq:inclusion}) are manifestly true. In applying~(\ref{eq:saturation}), it suffices to consider $ABCDE$ because it contains all RHS terms. Thus, when (\ref{eq:5party}) holds with equality then: \begin{equation}
E(ABCDE) = E(ABC) \cup E(BCD) \cup E(CDE) \cup E(DEA) \cup E(EAB)
\label{eq:5partyresultO}
\end{equation}

\begin{figure}[t]
    \centering
    \includegraphics[width=0.41\linewidth]{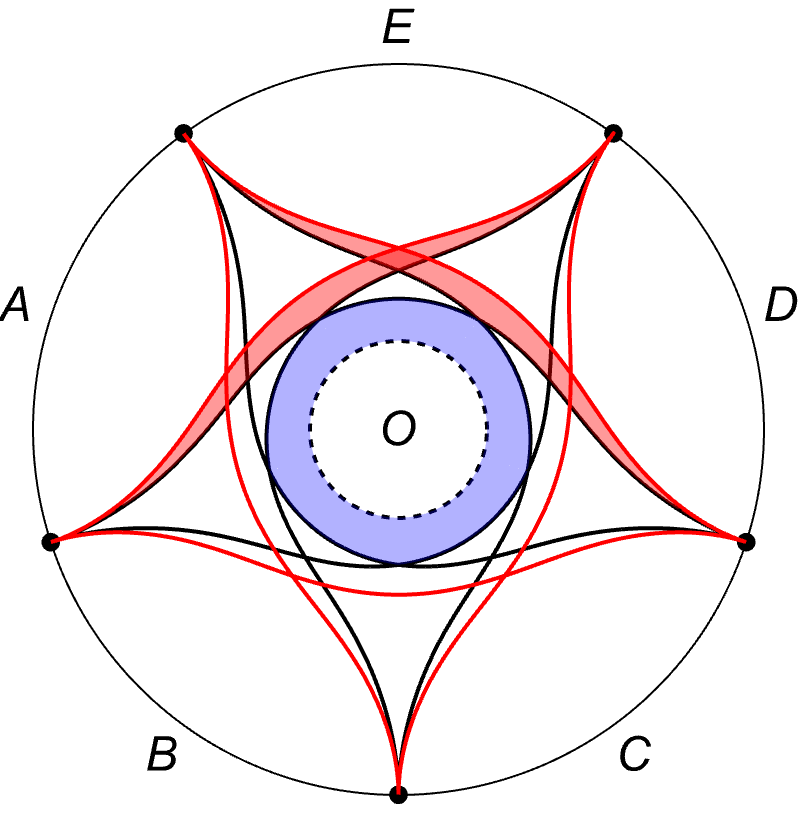}
    \caption{Bulk regions, which become empty when inequality~(\ref{eq:5party}) is saturated; (\ref{eq:5partyresultO}) is shown in blue and (\ref{eq:5partyresultE}) is shown in red. These regions lie deeper in the bulk (closer to the purifier $O$ / $E$) than $\cup_i E(X_i)$ does. Thus, if the inequality is saturated then the more `infrared' regions fail to exceed the depth reached by the `ultraviolet' regions.}
    \label{fig:dihedral_core}
\end{figure}

To see the effect of changing the purifier, introduce a purification $O$ and rewrite inequality~(\ref{eq:5party}) such that no term explicitly mentions $E$:
\begin{align}
& S(ABC) + S(BCD) + S(ABO) + S(BCO) + S(CDO) \label{eq:5party2} \\
\geq\,\,  & S(AB) + S(BC) + S(CD) + S(ABCO) + S(BCDO) + S(O) \nonumber
\end{align}
Now the validity of claim~(\ref{eq:inclusion}) relies on two regions: $ABCO$ and $BCDO$. Thus, when (\ref{eq:5party}) is saturated we have:
\begin{equation}
E(ABCO) \cup E(BCDO) = E(ABC) \cup E(BCD) \cup E(ABO) \cup E(BCO) \cup E(CDO)
\label{eq:5partyresultE}
\end{equation}
Figure~\ref{fig:dihedral_core} illustrates results~(\ref{eq:5partyresultO}) and (\ref{eq:5partyresultE}) using consecutively labeled intervals on the boundary of a static BTZ geometry \cite{Banados:1992wn}. The figure manifests the main qualitative point of the paper: {\bf When an inequality is saturated, the more `infrared' entanglement wedges fail to reach an incremental depth in the bulk.}

In fact, by exploiting details of contraction maps it is possible to prove that the saturation of (\ref{eq:5party}) implies these stronger relationships:
\begin{align}
E(ABCO) & = E(ABC) \cup E(ABO) \cup E(BCO) \label{eq:extra1} \\
E(BCDO) & = E(BCD) \cup E(BCO) \cup E(CDO) \label{eq:extra2}
\end{align}
This is the simplest example of a refinement of result~(\ref{eq:saturation}), which we advertised in the Introduction. We derive (\ref{eq:extra1}-\ref{eq:extra2}) and many similar results in the next section.

\section{A refinement}
\label{sec:main}
In Reference~\cite{Czech:2025jnw}, Wang and we explained that contraction proofs of holographic entropy inequalities strongly constrain what happens when the said inequalities are saturated. When those results are applied directly to inequality~(\ref{eq:schematic}), they are distinct and logically independent from claims~(\ref{eq:lemma}-\ref{eq:saturation}). However, a trick unifies the two approaches. If, instead of applying \cite{Czech:2025jnw} to inequality ${\rm LHS} \geq {\rm RHS}$, one applies it to
\begin{equation}
{\rm LHS} + S(Y_j) \geq {\rm RHS} + S(Y_j),
\label{redundant}
\end{equation}
claim~(\ref{eq:saturation}) follows \emph{along with} additional facts. In some cases, those additional facts significantly strengthen result~(\ref{eq:saturation}). 

In this section---and only here---we assume knowledge of proofs by contraction and of the results derived in Reference~\cite{Czech:2025jnw}. To make the paper self-contained, we review proofs by contraction in Appendix~\ref{sec:contraction} and the findings of \cite{Czech:2025jnw} in Appendix~\ref{sec:oldresult}. In order to establish common ground, we list below the necessary facts about contraction maps and implications of saturation---without justification. All facts highlighted below are justified in the Appendices.

\subsection{Key facts}
\label{sec:facts}
\paragraph{Proofs by contraction} When proving an inequality~(\ref{eq:schematic}), we consider binary strings $x$ whose components $x_i = 0~{\rm or}~1$ correspond to LHS terms $S(X_i)$. The method of proof is to construct a map $x \to f(x)$, where $f(x)$ are binary strings whose components $f(x)_j$ correspond to RHS terms $S(Y_j)$. 

Suppose there exists a map $f$, which satisfies two conditions:
\begin{itemize}
\item It is a contraction, viz.~(\ref{defcontraction}) in Appendix~\ref{sec:contraction}.
\item It satisfies inequality-dependent boundary conditions~(\ref{defbc}).
\end{itemize}
Then inequality~(\ref{eq:schematic}) is proven.

For quick reference, we mention that the boundary condition 
\begin{equation}
f(00\ldots 00) = 00\ldots 00
\label{bcpurifier}
\end{equation}
is imposed on contraction proofs of all inequalities written in a balanced form. This boundary condition is canonically associated with the purifier region.

\paragraph{Meaning of the contraction map}
We assume a time reversal-symmetric bulk spacetime and work on the static or time reversal-symmetric spatial slice. To each bit string $x$ we associate a subregion $W(x)$ of the bulk slice:
\begin{equation}
W(x) = \bigcap_i \Big( E(X_i)~{\rm or}~\overline{E(X_i)}\Big)
\end{equation}
where $E(X_i)$ is taken if $x_i = 1$ and $\overline{E(X_i)}$ if $x_i = 0$. From the viewpoint of holographic erasure correction, bit strings $x$ characterize the access structure to logical information localized in $W(x)$. Note that for some bit strings $x$, entanglement wedge nesting~(\ref{ewn1}-\ref{ewn2}) may directly imply $W(x) = \emptyset$. The conditions under which this happens are~(\ref{eq:xinconsistent}). Such bit strings are removed from the domain of the contraction map.

The point of the contraction map is to assemble spatial slices of candidate entanglement wedges of RHS regions according to:
\begin{equation}
E^{\rm can}(Y_j) = \cup_{x | x_j = 1} W(x)
\label{eq:defcan}
\end{equation}

\paragraph{Nesting violations}
If the candidate wedges~(\ref{eq:defcan}) were physical, the following circumstances would be inconsistent with entanglement wedge nesting:
\begin{equation}
\begin{array}{lp{5mm}l}
X_i \supset Y_j && {\rm and}~~~x_{i} = 0~~~{\rm and}~~~f(x)_j=1 \\
X_i \subset Y_j &&  {\rm and}~~~x_{i} = 1~~~{\rm and}~~~f(x)_j = 0 \\
X_i \cap Y_j = \emptyset &&  {\rm and}~~~x_{i} = 1~~~{\rm and}~~~f(x)_j = 1
\end{array}
\label{violations} 
\end{equation}
In particular, including $W(x)$ in $E(Y_j)$ if $f(x)_j = 1$, or excluding $W(x)$ from $E(Y_j)$ if $f(x)_j = 0$, would contradict~(\ref{ewn1}-\ref{ewn2}). Then the only way to have $E^{\rm can}(Y_j) = E(Y_j)$ is if $W(x)$ is empty.

\paragraph{Saturation implies that nesting-violating $W(x)$ are empty \cite{Czech:2025jnw}} If inequality~(\ref{eq:schematic}) is saturated (holds with equality) then [the spatial slices of] the candidate wedges \emph{are} indeed [slices of] the physical wedges. Therefore, if ${\rm LHS} = {\rm RHS}$ and any of circumstances~(\ref{violations}) holds then $W(x) = \emptyset$.

\paragraph{Roadmap}
In this section, we apply the aforementioned facts to the tautologically extended inequality
\begin{equation}
{\rm LHS} + S(Y_j) \geq {\rm RHS} + S(Y_j),
\tag{\ref{redundant}}
\end{equation}
which is saturated if and only if ${\rm LHS} \geq {\rm RHS}$ is. We shall find that this implies result~(\ref{eq:saturation}) and a number of other consequences. In some cases, those additional implications of saturation strengthen and refine claim~(\ref{eq:saturation}).

\subsection{Example: Subadditivity}
\label{sec:sa}
To get a feeling of how this works, let us re-examine the subadditivity of entanglement entropy. We shall rediscover the finding from the Introduction
\begin{equation}
S(A) + S(B) = S(AB) \quad \Longrightarrow \quad E(AB) = E(A) \cup E(B),
\label{eq:saresult}
\end{equation}
which is an instance of claim~(\ref{eq:saturation}).

The contraction $x \to f(x)$ that proves subadditivity is as follows:
\begin{table}[H]
\centering
\begin{tabular}{l cc p{0.2cm} c}
  \hline
  & \multicolumn{2}{c}{$x$} & & \multicolumn{1}{c}{$f({x})$} \\
  \cline{2-3} \cline{5-5}
      & ~$A$~ & ~$B$~  & & ~$AB$~ \\
$O$~~ & 0 & 0 & & 0   \\
$A$~~ & 1 & 0 & & 1   \\
$B$~~ & 0 & 1 & & 1   \\
  \rowcolor{gray!20} 
   ~~ & 1 & 1 & & - \\
  \hline
\end{tabular}
\caption{The contraction, which proves subadditivity. In the leftmost column we mark boundary conditions $(x^O, f^O)$ etc. Row $x=11$ is greyed out because it is automatically empty; viz.~(\ref{eq:xinconsistent}).}
\label{tab:contraction_sa}
\end{table}
\noindent
The $x=11$ row is greyed out because it refers to the automatically empty region $W(11) = E(A) \cap E(B)$. We observed that such bit strings---characterized in equations~(\ref{eq:xinconsistent})---are removed from the domain of the contraction \cite{Avis:2021xnz,lhsconflicts}. In what follows we do not write out such bit strings in contraction tables except when emphasizing that they are unphysical.

Now form inequality~(\ref{redundant}) by adding $S(AB) \geq S(AB)$ to subadditivity:
\begin{equation}
S(A) + S(B) + S(AB) \geq S(AB) + S(AB)
\label{sa:redundant}
\end{equation}
Note that (\ref{sa:redundant}) is saturated if and only if subadditivity is. The tautology $S(AB) \geq S(AB)$ is `proven' by the contraction map $f_t$:
\begin{table}[H]
\centering
\begin{tabular}{r c p{0.2cm} c}
  \hline
  & \multicolumn{1}{c}{$x_t$} & & \multicolumn{1}{c}{$f_t({x_t})$} \\
  \cline{2-2} \cline{4-4}
      & ~$AB$~ & & ~$AB$~ \\
$O$~~ & 0 & & 0   \\
$A, B$~~ & 1 & & 1   \\
  \hline
\end{tabular}
\caption{The contraction that `proves' every tautology $S(Y_j) \geq S(Y_j)$. The subscripts under $x_t$ and $f_t$ stand for `tautology.'}
\label{tab:tautology}
\end{table}
\noindent
To apply the technique of Section~\ref{sec:facts} to inequality~(\ref{sa:redundant})---and to exhibit the holographic consequences of saturating subadditivity---we must write out a contraction proof of (\ref{sa:redundant}). This simply concatenates Table~\ref{tab:contraction_sa} and Table~\ref{tab:tautology} into a combined map $(x, x_t) \to (f(x), f_t(x_t))$:
\begin{table}[H]
\centering
\begin{tabular}{l ll @{}p{0.25cm} @{}c @{} p{0.8cm} @{} c @{} p{0.10cm} @{} c}
  \hline
  & \multicolumn{2}{c}{$x$} & & $x_t$ & & \multicolumn{1}{c}{$f(x)$} & & \multicolumn{1}{c}{$f_t(x_t)$} \\
  \cline{2-3} \cline{5-5} \cline{7-7} \cline{9-9}
            & $A$ & $B$~  & & $AB$            & & $AB$            & & $AB$ \\
$O$~~ &   0   &    0    & &    0                  & &     0               & & 0 \\
\rowcolor{gray!20} 
            &   1   &    0     & &    0                 & &      1              & & 0 \\
\rowcolor{gray!20} 
            &   0   &    1     & &    0                & &      1               & & 0 \\
            &   0   &    0    & & \framebox{1}  & & \framebox{0} & & 1 \\
$A$~~ &   1   &    0     & &    1                 & &      1              & & 1 \\
$B$~~ &   0   &    1     & &    1                 & &      1              & & 1 \\
  \hline
\end{tabular}
\caption{The contraction that proves~(\ref{sa:redundant}). The greyed out bit strings are removed due to conditions~(\ref{eq:xinconsistent}). The boxed entries are an instance of (\ref{violations}) so $W(001)$ is empty at saturation.}
\label{tab:contraction_red}
\end{table}
\noindent
In this contraction map, the rows with $x = 10, 01$ and $x_t = 0$ are instances of (\ref{eq:xinconsistent}). They are automatically empty by entanglement wedge nesting and can be removed from the contraction table altogether. 

On the other hand, the row $x = 00$ and $x_t = 1$ is physical. It represents the region:
\begin{equation}
W(001) = \overline{E(A)} \cap \overline{E(B)} \cap E(AB) = \overline{E(A) \cup E(B)} \cap E(AB)
\label{w001}
\end{equation}
The boxed entries in this row are an instance of (\ref{violations}) and therefore $W(001)$ is empty when (\ref{sa:redundant}) is saturated. Now $W(001) = \emptyset$ is equivalent to $E(AB) \subseteq E(A) \cup E(B)$. Since the opposite inclusion ($\supseteq$) holds by wedge nesting, the two must be equal and we have recovered~(\ref{eq:saresult}).

\begin{figure}[t]
\begin{center}
\raisebox{0\linewidth}{
\includegraphics[width=0.28\linewidth]{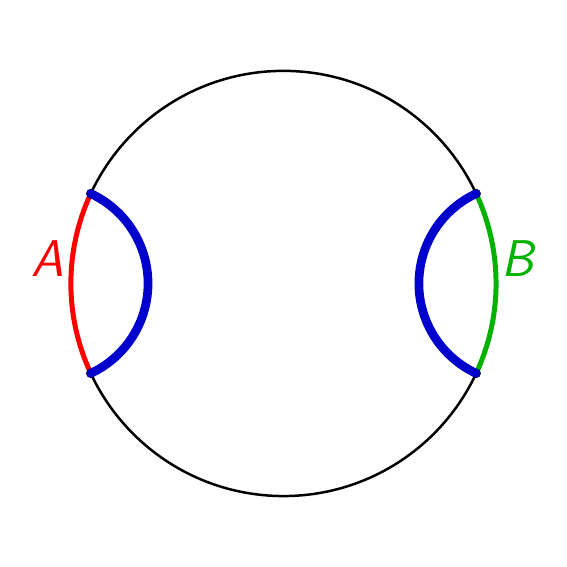}
}
\hspace*{0.01\columnwidth}
\includegraphics[width=0.28\linewidth]{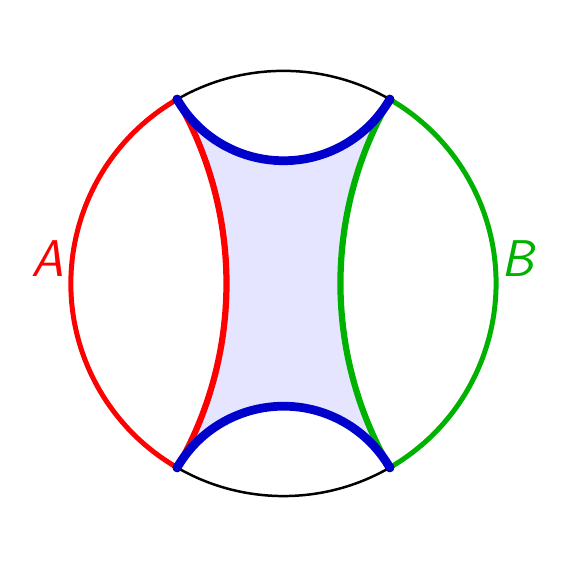}
\hspace*{0.01\columnwidth}
\includegraphics[width=0.28\linewidth]{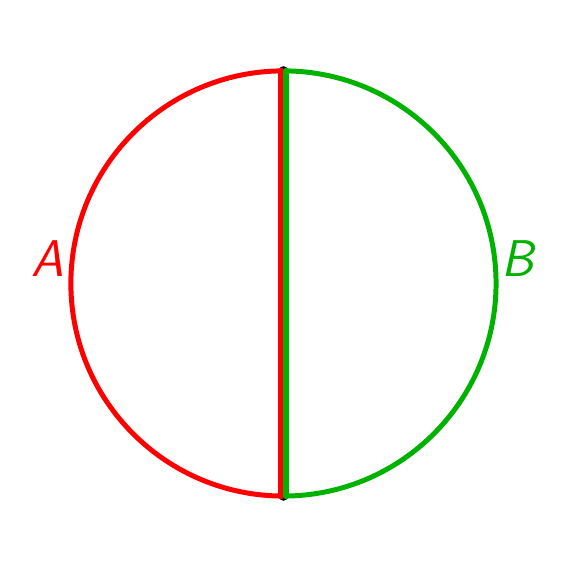}
\caption{Left: If $S(A) + S(B) = S(AB)$ then region~(\ref{w001}) is empty. Middle: If region~(\ref{w001}) is nonempty then $S(A) + S(B) > S(AB)$. Right: The converse of~(\ref{eq:saresult}) is not true.}
\label{fig:sat_sa}
\end{center}
\end{figure}

This conclusion is easy to see in examples, see Figure~\ref{fig:sat_sa}. Note that the reverse implication does not hold. A pure state on $AB$ also has ${\rm EW}(AB) = {\rm EW}(A) \cup {\rm EW}(B)$ yet $S(AB) = 0$.

\subsection{General argument}
\label{sec:generalities}
We extract a pattern from the previous example and re-derive claim~(\ref{eq:saturation}). We work with a general inequality ${\rm LHS} \geq {\rm RHS}$ and apply the facts from Section~\ref{sec:facts} to the redundant inequality ${\rm LHS} + S(Y_j) \geq {\rm RHS} + S(Y_j)$, where $Y_j$ is one of the regions on the RHS. 

The first step is to form a contraction proof of ${\rm LHS} + S(Y_j) \geq {\rm RHS} + S(Y_j)$. This is done by concatenating the contraction map $x \to f(x)$ that proves ${\rm LHS} \geq {\rm RHS}$ with the contraction that proves $S(Y_j) \geq S(Y_j)$, which is given in Table~\ref{tab:tautology}. We thus end up with a duplicated contraction table:
\begin{table}[H]
\centering
\begin{tabular}{c p{0.01cm} c p{0.3cm} c p{0.01cm} c}
  \hline
LHS & & $Y_j$ & & RHS & & $Y_j$ \\
  \cline{1-1} \cline{3-3} \cline{5-5} \cline{7-7}
$x$~  & & $0$            & & $f(x)$            & & 0 \\
$x$~  & & $1$            & & $f(x)$            & & 1 \\
  \hline
\end{tabular}
\caption{Contraction map for inequality ${\rm LHS} + S(Y_j) \geq {\rm RHS} + S(Y_j)$.}
\label{tab:genred}
\end{table}
\noindent
Note that the $j^{\rm th}$ bit of $f(x)$---which we denote $f(x)_j$---also describes region $Y_j$ as part of the RHS of the original inequality. Therefore, \emph{for every} $x \in \{0,1\}^L$, one of the rows in this duplicated table necessarily features a contraction violation~(\ref{violations}). This activates the conclusions of Section~\ref{sec:facts} and implies:
\begin{itemize}
\item If $f(x)_j = 0$ then the lower row in Table~\ref{tab:genred} violates nesting. We conclude:
\begin{equation}
{\rm LHS} = {\rm RHS} \qquad \Longrightarrow \qquad W(x) \cap E(Y_j) = \emptyset
\label{fj0}
\end{equation}
\item If $f(x)_j = 1$ then the upper row in Table~\ref{tab:genred} violates nesting. We conclude:
\begin{equation}
{\rm LHS} = {\rm RHS} \qquad \Longrightarrow \qquad W(x) \cap \overline{E(Y_j)} = \emptyset
\label{fj1}
\end{equation} 
\end{itemize}
We remark that the combined bit string $(x, 0)$ or $(x,1)$ may fall under conditions~(\ref{eq:xinconsistent}). In such cases, implications~(\ref{fj0}) and (\ref{fj1}) are trivial: the consequent holds by virtue of entanglement wedge nesting alone, irrespective of the saturation of the inequality.

\paragraph{Recovering result~(\ref{eq:saturation})}
Apply (\ref{fj0}) to the boundary condition $f(00\ldots 00) = 00\ldots 00$ on any RHS term $Y_j$. We find that if ${\rm LHS} = {\rm RHS}$ then:
\begin{equation}
\Big( \cap_i \overline{E(X_i)}\Big) \cap E(Y_j) = \emptyset
\label{eq:firstline}
\end{equation}
This can be rewritten in the more illuminating form:
\begin{equation}
E(Y_j) \subseteq \cup_i E(X_i)
\label{eq:singleinc}
\end{equation}
We now take the union of (\ref{eq:singleinc}) over all $Y_j$:
\begin{equation}
\cup_j E(Y_j) \subseteq \cup_i E(X_i)
\label{eq:allinc}
\end{equation}
Claim~(\ref{eq:inclusion}) says that the reverse inclusion ($\supseteq$) holds generally. Altogether, we find that if ${\rm LHS} = {\rm RHS}$ then $\cup_j E(Y_j) = \cup_i E(X_i)$, which is claim~(\ref{eq:saturation}) from Section~\ref{sec:rg}.

\paragraph{Example: The five-party dihedral inequality}
For illustration, consider the five-party dihedral inequality written in form~(\ref{eq:5party2}). Observe that claim~(\ref{eq:lemma}) picks out $ABCO$ and $BCDO$ as the `largest' RHS terms, in the sense that one or both of them contains every LHS term.

Extend (\ref{eq:5party2}) by the tautology~$S(ABCO) \geq S(ABCO)$. Applying result~(\ref{fj0}) to the boundary condition $f(00000) = 000000$, we find that the saturation of (\ref{eq:5party2}) implies:
\begin{equation}
E(ABCO) \subseteq E(ABC) \cup E(BCD) \cup E(ABO) \cup E(BCO) \cup E(CDO) = \cup_i E(X_i)
\end{equation}
Extending (\ref{eq:5party2}) by $S(BCDO) \geq S(BCDO)$ produces an analogous result. Taking the union of the two, we have:
\begin{equation}
E(ABCO) \cup E(BCDO) \subseteq \cup_i E(X_i)
\end{equation}
Combining this with result~(\ref{eq:inclusion}) we arrive at conclusion~(\ref{eq:5partyresultE}).

We remark that it suffices to run the argument in lines~(\ref{eq:firstline}-\ref{eq:allinc}) over those $Y_j$, which arise in the application of result~(\ref{eq:lemma}). In the language of the Renormalization Group, they make the RHS `more infrared' than the LHS.

\subsection{The refinement}
\label{sec:applications}
By applying conclusions~(\ref{fj0}-\ref{fj1}) to bit strings $x$ other than $x = 00\ldots 00$, it is sometimes possible to significantly strengthen result~(\ref{eq:saturation}). In this subsection we exhibit this refinement as it arises in the infinite families of holographic entropy inequalities \cite{Czech:2023xed}.

\subsubsection{Dihedral inequalities}
\label{sec:dihedrals}
Consider an odd number $m$ of regions $A_i$, which are cyclically indexed, i.e. $A_{m+1} \equiv A_1$ etc. Define:
\begin{align}
A_i^- & = A_i \,A_{i+1} \ldots A_{i+(m-3)/2} \label{defap} \\
A_i^+ & = A_i \,A_{i+1} \ldots A_{i+(m-1)/2} \label{defam}
\end{align}
The dihedral inequalities \cite{Bao:2015bfa} take the form:
\begin{equation}
\sum_{i=1}^m S(A_i^+) \,\geq\, \sum_{i=1}^m S(A_i^-) + S(A_1 A_2 \ldots A_m)
\label{dihedral}
\end{equation}
The monogamy of mutual information~(\ref{eq:mmi}) is the instance with $m=3$. We have also encountered the $m=5$ instance in inequality~(\ref{eq:5party}). 

\paragraph{Canonical purifier}
Inequality~(\ref{dihedral}) has a unique `most infrared' term $S(A_1 A_2 \ldots A_m)$. Therefore, if it is saturated then:
\begin{equation}
E(A_1 A_2 \ldots A_m) = \cup_i E(A_i^+)
\label{cyclic1term}
\end{equation}
Conclusions~(\ref{mmiresult1}) and (\ref{eq:5partyresultO}) are special instances of this. We have not yet \emph{refined} our prior conclusions.

\paragraph{A different purifier}
Now let us rewrite~(\ref{dihedral}) setting $A_m$ as the purifier:
\begin{equation}
\sum_{i=1}^{(m+1)/2} S(A_i^- O) + \sum_{i=1}^{(m-1)/2} S(A_i^+)
\,\geq
\sum_{i=1}^{(m+1)/2} S(A_i^-) + \sum_{i=1}^{(m-1)/2} S(A_i^+ O) + S(O)
\label{dihedral2}
\end{equation}
The contraction map that proves (\ref{dihedral2}) has the following interesting properties:
\begin{table}[H]
\centering
\begin{tabular}{cccccc p{0.30cm} ccc}
  \hline
  \multicolumn{6}{c}{$x$} & & \multicolumn{3}{c}{$f(x)$} \\
  \cline{1-6} \cline{8-10}
   $A_{1}^+$ & $A_2^+$ & $A_1^- O$ & $A_2^- O$ & $A_3^- O$ & {\rm others} 
      & & $A_1^+ O$ & $A_2^+ O$ & {\rm others}$\phantom{\Big|}$\\[3pt]
       0 & 0 &  0 & 0 & 0 & 0
      & & 0 & 0 & 0 \\[3pt]
   0 & {\rm any} & 0 & 0 & {\rm any} & {\rm any}
      & & \framebox{0} & $\times$ & $\times$ \\[2pt]
    {\rm any} & 0 & {\rm any} & 0 & 0 & {\rm any}
      & & $\times$ & \framebox{0} & $\times$ \\[2pt]
  \hline
\end{tabular}
\caption{Part of the contraction map, which proves~(\ref{dihedral2}). Applications of (\ref{fj0}), which refine result~(\ref{eq:saturation}), are highlighted in boxes.}
\label{tab:dihedralewn}
\end{table}
\noindent
In particular, the highlighted values of $f(x)_j$ persist for \emph{all} choices of bits $x_i$, which are left undetermined (`any') in Table~\ref{tab:dihedralewn}. These properties were proven for all values of $m$ in \cite{Bao:2015bfa}. It is amusing to confirm them with the explicit contraction map for $m=5$, which is found in Appendix~B of \cite{Bao:2015bfa}.

Apply~(\ref{fj0}) to bit strings $x$ in the second row of Table~\ref{tab:dihedralewn}. We find that if (\ref{dihedral2}) is saturated then
\begin{equation}
E(A_1^+ O) \cap \overline{E(A_1^+)} \cap \overline{E(A_1^- O)} \cap \overline{E(A_2^- O)} \cap \ldots 
= \emptyset
\end{equation}
where $\ldots$ denotes \emph{any} overlap of the remaining $E(X_i)$ or $\overline{E(X_i)}$. Taking the union over all such statements, we find simply:
\begin{equation}
E(A_1^+ O) \subseteq 
\overline{\overline{E(A_1^+)} \cap \overline{E(A_1^- O)} \cap \overline{E(A_2^- O)}\phantom{\Big|}} 
= E(A_1^+) \cup E(A_1^- O) \cup E(A_2^- O)
\end{equation}
As entanglement wedge nesting guarantees the opposite inclusion ($\supseteq$), we recognize that saturating (\ref{dihedral}) implies:
\begin{equation}
E(A_1^+ O) = E(A_1^+) \cup E(A_1^- O) \cup E(A_2^- O)
\end{equation}
The same argument applied to the third row of Table~\ref{tab:dihedralewn} gives:
\begin{equation}
E(A_2^+ O) = E(A_2^+) \cup E(A_2^- O) \cup E(A_3^- O)
\end{equation}
These two conclusions refine statement~(\ref{eq:saturation}) when $A_m$ is the purifier. They generalize conclusions (\ref{eq:extra1}-\ref{eq:extra2}) to dihedral inequalities, which feature an arbitrary odd number of regions $A_i$. 

After changing purifiers, we find that the saturation of (\ref{dihedral}) implies for all $k$:
\begin{equation}
E(A_k^+ O) = E(A_k^+) \cup E(A_k^- O) \cup E(A_{k+1}^- O)
\label{cyclicgen}
\end{equation}

\subsubsection{Toric inequalities}
\label{sec:toric}
In \cite{Czech:2022fzb, Czech:2023xed, Czech:2024rco}, the following generalization of the dihedral inequalities was proposed and proven:
\begin{equation}
\sum_{i=1}^m \sum_{j=1}^n S(A_i^+ B_j^-) \,\geq\, \sum_{i=1}^m \sum_{j=1}^n S(A_i^- B_j^-) + S(A_1 A_2 \ldots A_m)
\label{eq:toric}
\end{equation}
The notation follows definitions~(\ref{defap}-\ref{defam}). Presented in this form, inequality~(\ref{eq:toric}) applies to \emph{pure states} on $A_1 \ldots A_m B_1 \ldots B_n$. Note that presentation~(\ref{eq:toric}) is not balanced and does not designate any particular region as the purifier. Nevertheless, inequalities~(\ref{eq:toric}) are superbalanced and any region $A_i$ or $B_j$ \emph{can} be cast in the role of the purifier.

The dihedral inequalities~(\ref{dihedral}) are the special case where $n=1$, but in what follows we concentrate \emph{exclusively} on $n\geq 3$ because the analyses differ in important details.

\paragraph{Refined consequences of saturation}
Let us skip the steps of selecting a purifier and rewriting the inequality in a balanced form. Instead, work with a contraction map that applies directly to presentation~(\ref{eq:toric}). Note that at this point the bit string $x = 00\ldots 00$ no longer plays any special role. However, we shall eventually recognize the outcome of this analysis as a refinement of result~(\ref{eq:saturation}) along the lines of (\ref{eq:refined}).

As shown in \cite{Czech:2023xed}, contraction proofs of inequality~(\ref{eq:toric}) exhibit the following pattern:
\begin{table}[H]
\centering
\begin{tabular}{ccccc p{0.30cm} cc}
  \hline
  \multicolumn{5}{c}{$x$} & & \multicolumn{2}{c}{$f(x)$} \\
  \cline{1-5} \cline{7-8}
     \raisebox{17mm}{$A_i^+ B_j^-$} 
  & \raisebox{17mm}{$A_i^+ B_{j+1}^-$} 
  & \rotatebox{90}{~$A_{i+(m-1)/2}^+ B_{j+(n+1)/2}^-$~~} 
  & \rotatebox{90}{~$A_{i+(m+1)/2}^+ B_{j+(n+1)/2}^-$~~} 
  & \raisebox{17mm}{\rm others}
  & 
   & \raisebox{17mm}{$A_{i+(m+1)/2}^-B_{j+(n+1)/2}^-$} 
  & \raisebox{17mm}{\rm others}\\
  0 & 0 & ~~~1~~~ & ~~~1~~~ & {\rm any}
      & & ~\framebox{1}~ & $\times$ \\[1pt]
  \hline
\end{tabular}
\caption{Part of the contraction map, which proves~(\ref{eq:toric}). The box highlights the application of (\ref{fj1}), which refines result~(\ref{eq:saturation}).}
\label{tab:toricewn}
\end{table}
\noindent
Applying~(\ref{fj1}) to bit strings $x$ displayed in Table~\ref{tab:toricewn}, we find this consequence of the saturation of (\ref{eq:toric}):
\begin{align}
& \overline{E(A_i^+B_j^-)} \cap \overline{E(A_i^+B_{j+1}^-)} 
\cap E(A_{i+(m-1)/2}^+ B_{j+(n+1)/2}^-)
\cap E(A_{i+(m+1)/2}^+ B_{j+(n+1)/2}^-)
\cap \ldots \nonumber \\
& ~~~~~~\cap \overline{E(A_{i+(m+1)/2}^-B_{j+(n+1)/2}^-)} = \emptyset
\label{eq:rawconcl}
\end{align}
As in the treatment of the dihedral inequalities, let us take the union over all unspecified regions $E(X_i)$ or $\overline{E(X_i)}$. Turning the overlap of complements into the complement of the union and using $\overline{E(X)} = E(\overline{X})$ where appropriate, we obtain:
\begin{equation}
E(A_i^+ B_j^+) \subseteq E(A_i^+ B_j^-) \cup E(A_i^+ B_{j+1}^-) \cup E(A_i^- B_j^+) \cup E(A_{i+1}^- B_j^+)
\end{equation}
Since the reverse containment relationship holds by entanglement wedge nesting, we conclude that when (\ref{eq:toric}) is saturated:
\begin{equation}
E(A_i^+ B_j^+) = E(A_i^+ B_j^-) \cup E(A_i^+ B_{j+1}^-) \cup E(A_i^- B_j^+) \cup E(A_{i+1}^- B_j^+)
\label{eq:toricrefined}
\end{equation}
for all $i$ and $j$. Conclusion~(\ref{eq:toricrefined}) is a refinement of claim~(\ref{eq:saturation}) for every choice of purifier not contained in $A_i^+ B_j^+$. 

\paragraph{Note} In References~\cite{Czech:2023xed, Czech:2025tds}, we introduced a graphical representation of bit strings $x$, which involves nonintersecting loops living on an $n \times m$-square tessellation of a torus. The rows highlighted in Table~\ref{tab:toricewn}---and the refined statement~(\ref{eq:toricrefined})---arise from those bit strings, which are represented by \emph{minimal} loops.

\subsubsection{Projective plane inequalities}
Consider $m$ cyclically ordered regions $A_i$ and $B_i$ with $A_{i+m} \equiv A_i$ and $B_{i+m} \equiv B_i$. To write down the projective plane inequalities, we introduce the notation
\begin{equation}
A_i^{(j)} = A_i A_{i+1} \ldots A_{i+j-1}
\end{equation}
and likewise for $B_i^{(j)}$. Referring back to definitions~(\ref{defap}-\ref{defam}), note that if $m$ is odd then $A_i^\pm = A_i^{((m\pm 1)/2)}$. We assume that the global state on $A_1 \ldots A_m B_1 \ldots B_m$ is pure. 

The projective plane inequalities take the form \cite{Czech:2023xed, Czech:2024rco}:
\begin{align} 
\frac{1}{2} 
\sum_{i=1}^{m} \sum_{j = 1}^{m-1}&
\Bigg(S\left(A_i^{(j)} B_{i+j-1}^{(m-j)}\right) + S\left(A_i^{(j)} B_{i+j}^{(m-j)}\right) \Bigg) 
\, + (m-1)\, S\Big(A_1A_2\ldots A_m\Big) \nonumber \\
& \geq  
\sum_{i=1}^{m} \sum_{j=1}^{m} S\left(A_i^{(j-1)} B_{i+j-1}^{(m-j)}\right)
\label{rp2ineqs}
\end{align}
Here $m$ can be either odd or even. On the LHS, each independent term in fact appears with unit factor because every $S(Y)$ is accompanied by $S(\overline{Y})$. We present this form of the inequality, with the ostensible factor of 1/2, because it preserves the most symmetry. As a notable special case, inequality~(\ref{rp2ineqs}) for $m = 2$ coincides with the monogamy of mutual information~(\ref{eq:mmi}). 

\begin{table}[t]
\centering
\begin{tabular}{ccccc p{0.20cm} c}
  \hline
  \multicolumn{5}{c}{$x$} & & \multicolumn{1}{c}{$f(x)$} \\
    \cline{1-5} \cline{7-7}
  \\[-9pt]
     {$A_i B_i^{(m-1)}$} 
  & {$A_i^{(m-1)} B_{i-1}$} 
  & \multicolumn{2}{c}{$A_1 A_2 \ldots A_m$} 
  & {\rm others}
  & 
   & {$B_{i}^{(m-1)}$}\\[3pt]
  1 & 0 & \multicolumn{2}{c}{0} & {\rm any}
      & & ~\framebox{1}~\\[2pt]
  \cline{1-5} \cline{7-7}
     \raisebox{13mm}{$A_i^{(j)} B_{i+j-1}^{(m-j)}$} 
  & ~~\rotatebox{90}{~~~$A_{i+j-1}^{(m-j)} B_{i-1}^{(j)}$}~~
  & ~~\rotatebox{90}{~$A_i^{(j-1)} B_{i+j-2}^{(m-j+1)}$~}~~
  & \raisebox{13mm}{~$A_{i+j-1}^{(m-j+1)} B_i^{(j-1)}$} 
  & \raisebox{13mm}{\rm others}
  & 
   & \raisebox{13mm}{$A_i^{(j-1)} B_{i+j-1}^{(m-j)}$}\\[1pt]
  1 & 0 & 1 & 0 & {\rm any}
      & & ~\framebox{1}~\\[3pt]
  \cline{1-5} \cline{7-7}
  \\[-9pt]
      \multicolumn{2}{c}{$A_1 A_2 \ldots A_m$} 
  & {$A_i^{(m-1)} B_{i-2}$} 
  &  {$A_{i-1} B_i^{(m-1)}$}
  & {\rm others}
  & 
   & {$A_i^{(m-1)}$}\\[3pt]
  \multicolumn{2}{c}{1} & 1 & 0 & {\rm any}
      & & ~\framebox{1}~\\[2pt]
        \hline
\end{tabular}
\caption{Part of the contraction map, which proves~(\ref{rp2ineqs}). Boxes highlight applications of (\ref{fj1}), which refine result~(\ref{eq:saturation}).}
\label{tab:rp2ewn}
\end{table}

\paragraph{Refined consequences of saturation} 
Our steps mirror Section~\ref{sec:toric}. In contraction maps given in \cite{Czech:2023xed}, certain values of $f(x)_j$ persist for \emph{all} $x$-bits, which appear as `any' in Table~\ref{tab:rp2ewn}. In each row, taking the union over all implications of saturation indicated in (\ref{fj1}) gives: 
\begin{align}
E(A_i B_i^{(m-1)}) \cap \overline{E(A_i^{(m-1)} B_{i-1}}) \cap \overline{E(A_1 A_2 \ldots A_m)} 
\cap \overline{E(B_i^{(m-1)})}
& = \emptyset\phantom{\Bigg|}
\label{rpresult1}
\\
E(A_i^{(j)} B_{i+j-1}^{(m-j)}) \cap \overline{E(A_{i+j-1}^{(m-j)} B_{i-1}^{(j)})} 
\cap E(A_i^{(j-1)} B_{i+j-2}^{(m-j+1)}) \nonumber \\
\cap\, \overline{E(A_{i+j-1}^{(m-j+1)} B_i^{(j-1)})} \cap \overline{E(A_i^{(j-1)} B_{i+j-1}^{(m-j)})}
& = \emptyset
\label{rpresult2}
\\
E(A_i^{(m-1)} B_{i-2}) \cap \overline{E(A_{i-1} B_i^{(m-1)})} \cap E(A_1 A_2 \ldots A_m)
\cap \overline{E(A_i^{(m-1)})}
& = \emptyset\phantom{\Bigg|}
\label{rpresult3}
\end{align}
After following the same steps that took us from (\ref{eq:rawconcl}) to (\ref{eq:toricrefined}), we find that the saturation of (\ref{rp2ineqs}) implies:
\begin{align}
E(A_{\rm all} B_{i-1}) & = E(A_{i}^{(m-1)} B_{i-1}) \cup E(A_{i+1}^{(m-1)} B_{i-1}) \cup E(A_{\rm all})
\label{rpref1}
\\
E(A_{i+j-1}^{(m-j+1)} B_{i-1}^{(j)}) & = 
\nonumber \\
E(A_{i+j-1}^{(m-j)} B_{i-1}^{(j)})
& \cup E(A_{i+j}^{(m-j)} B_{i-1}^{(j)}) 
\cup E(A_{i+j-1}^{(m-j+1)} B_{i-1}^{(j-1)})
\cup E(A_{i+j-1}^{(m-j+1)} B_i^{(j-1)})
\label{rpref2} \\
E(A_{i-1} B_{\rm all}) & = 
E(A_{i-1} B_{i-1}^{(m-1)}) \cup E(A_{i-1} B_i^{(m-1)}) \cup E(B_{\rm all})
\label{rpref3}
\end{align}
Here $A_{\rm all} = A_1 A_2 \ldots A_m$ and likewise for $B_{\rm all}$. These statements refine result~(\ref{eq:saturation}) for every choice of purifier, which is not contained in the respective `infrared,' $(m+1)$-partite region. 

\paragraph{Note} As in the case of the toric inequalities (Section~\ref{sec:toric}), these results become more transparent when organized using the graphical formalism developed in \cite{Czech:2023xed, Czech:2025tds}. They again descend from bit strings $x$, which are represented by minimal loops.

\section{Discussion}
For a long time, holographic entropy inequalities have been an answer in search of a question. Previous works have pointed out applications to black hole evaporation \cite{Czech:2021rxe} and quantum erasure correction \cite{Czech:2025jnw} and, indeed, the material in Section~\ref{sec:main} continues and extends the latter line of thought. But the question posed by the Renormalization Group is general and compelling: 
\begin{itemize}
\item What boundary quantities are order parameters for the additional bulk depth, which is reached by entanglement wedges of infrared regions? 
\end{itemize}
The holographic entropy cone answers this question with statements (\ref{eq:inclusion}) and (\ref{eq:saturation}).

In our view, it does so in a rather clever way. The nonnegative quantities defined by the inequalities are UV-finite and therefore exclusively sensitive to physics away from the asymptotic boundary. Moreover, the notion of bulk depth is adjustable---as indeed it should be in a system described by a \emph{conformal} field theory. We explained in Section~\ref{sec:inclusion} that UV-finiteness---manifested by the inequalities with the property of superbalance \cite{He:2020xuo}---ensures that the relation between holographic RG and the inequalities accommodates every convention of what we designate as the CFT infrared.

The most important finding in this paper is that the saturation of a holographic entropy inequality implies a special structure among entanglement wedges: wedges of larger (infrared) regions, which should ordinarily reach deeper into the bulk than do wedges of smaller (ultraviolet) regions, fail to reach additional depth. By extrapolation, one may speculate that quantum states that violate holographic entropy inequalities can describe bulk geometries, which violate entanglement wedge nesting: ultraviolet wedges reaching deeper than infrared wedges? Engineering such a circumstance requires violating the null energy condition \cite{maximin} and amounts to constructing a traversable wormhole \cite{Gao:2016bin}. Can holographic entropy inequalities detect traversable wormholes? 

In Section~\ref{sec:main}, we showed that the saturation of inequalities in the infinite families entails more detailed consequences, which refine and improve result~(\ref{eq:saturation}). These consequences can be understood both as statements about holographic RG and about access structures of quantum erasure correcting codes \cite{Czech:2025jnw}. This highlights a relationship between holographic RG and erasure correction, which is obvious but seldom emphasized: running the Renormalization Group improves the protection of logical data. Holographic entropy inequalities endow this slogan with crisp meaning, which is manifested in statements like (\ref{cyclic1term}) and Figure~\ref{fig:dihedral_core}. 

\paragraph{Comments about the holographic entropy cone} Instead of its primitive inequalities (facets), the holographic entropy cone can also be described by its extreme rays---values of subsystem entropies, which saturate as many inequalities as possible. We have seen here that the saturation of every inequality implies \emph{many} statements about wedges failing to reach deeper than other wedges. It is interesting to understand the extreme rays in the same language. A first expectation is that they correspond to geometries, in which holographic RG `stops' almost everywhere and only one or a few combinations of larger regions contain independent infrared data. We plan to study this issue in upcoming work.

We believe that the RG-related interpretation of holographic entropy inequalities adds evidence in favor of the conjecture of \cite{Bao:2025sjn} that \emph{all} such inequalities can be proved by the contraction method. That is because our results rely strongly on contraction proofs. Assuming the connection with RG is not an accident of geometry, it would be strange to find an inequality, which maintains the relationship with the holographic Renormalization Group in spirit but not in its technical content. 

\paragraph{Toward generalizations} An obvious question concerns the issue of time dependence. The techniques we used---first and foremost, contraction maps---assume that all RT surfaces live on a common spatial slice of the bulk, which is manifestly false for HRT surfaces \cite{hrt, maximin}. Thus, our results---at least in their present form---are strongly anchored to time reversal symmetry. It is also clear that our assertions about spatial slices $E(Y)$ of entanglement wedges do not lift well to full entanglement wedges ${\rm EW}(Y)$, which are the domains of dependence of the $E(Y)$. For example, statement~(\ref{eq:saturation}) is not true if we substitute $E(Y) \to {\rm EW}(Y)$ for the simple reason that the domain of dependence of the union is not the union of the domains of dependence.

As emphasized in \cite{Grimaldi:2025jad}, only the issue of time dependence ties the holographic entropy cone to gravitational dynamics. Thinking along these lines has already proven remarkably fruitful for understanding the structure of holographic entropy inequalities. Thus, it is imperative to try to understand whether and how claim~(\ref{eq:saturation}) can be lifted to time-dependent settings.

Claims~(\ref{eq:lemma}) and (\ref{eq:inclusion}) hold under all circumstances, including time dependence and corrections due to bulk entanglement. But it is not clear if one should anticipate claim~(\ref{eq:saturation}) to have a generalization, which accommodates bulk entanglement. Areas of quantum extremal surfaces \cite{Engelhardt:2014gca} can famously violate holographic entropy inequalities. (Additional conditions under which they respect them were studied in \cite{sergioqes}.) The existence of such a generalization is perhaps the most stringent test of the conceptual significance of result~(\ref{eq:saturation}).

\acknowledgments
We thank Bowen Chen, Keiichiro Furuya, Yichen Feng, Matthew Headrick, Alex Maloney, Eva Silverstein, Leonard Susskind, Yixu Wang, Minjun Xie and Dachen Zhang for discussions. BC thanks the organizers of the workshops `Quantum Gravity, Holography, and Quantum Information' held at IIP Natal (Brazil) and `Discussions on Quantum Spacetime' held at LMSI (Mumbai, India) and acknowledges the hospitality of UC Berkeley and Stanford University, where part of this work was completed. This research was supported by an NSFC grant number 12042505 and a BJNSF grant under the Gao Cengci Rencai Zizhu program.
\medskip

\appendix
\section{Contraction maps}
\label{sec:contraction}
All known holographic inequalities have been proven using a technique called contraction map \cite{Bao:2015bfa}. For an inequality with the schematic structure~(\ref{eq:schematic}), one constructs a map $f: \{0,1\}^L \to \{0,1\}^R$, where $L$ and $R$ count LHS and RHS terms in the inequality. Here is how and why maps $f$ prove entropic inequalities:

\paragraph{Domain of $f$: Access structures}
The bit strings $x \in \{0,1\}^L$ in the domain of $f$ are associated with subregions $W(x)$ of the bulk time reversal-symmetric slice, where entanglement wedges of regions $X_i$ or $\overline{X_i}$ intersect. We associate $x_i = 1$, that is the $i^{\rm th}$ component of bit string $x$, with the equal time slice of the entanglement wedge $E(X_i)$ and $x_i = 0$ with the same for $E(\overline{X_i}) = \overline{E(X_i)}$:
\begin{equation}
W(x) = \bigcap_{i=1}^L E\big(X_i^{(x_i)}\big) \qquad {\rm where}~~X_i^{(x_i)} = 
\begin{cases} 
X_i & {\rm if}~x_i = 1 \\
\overline{X_i} & {\rm if}~x_i = 0
\end{cases}
\label{defwx}
\end{equation}
Observe that, by definition, the equal time cross section of the entanglement wedge of $X_i$ (respectively $\overline{X_i}$) is:
\begin{equation}
E(X_i) = \cup_{x|x_i=1} W(x) \qquad {\rm and} \qquad \overline{E(X_i)} = E(\overline{X_i}) = \cup_{x|x_i=0} W(x)
\label{ewtautology}
\end{equation}

In general, many regions $W(x)$ may be empty. When we view the bulk geometry as an erasure correcting code, definition~(\ref{defwx}) and subregion duality \cite{Czech:2012bh,Dong:2016eik} assert that the bulk degrees of freedom in $W(x)$ are recoverable from each $X_i^{(x_i)}$ and protected against the erasure of each $\overline{X_i^{(x_i)}}$. (Note that our definitions imply $\overline{X_i^{(0)}} = X_i$.) Therefore, each bit string $x$ characterizes an access structure, which the code realizes if and only if $W(x) \neq \emptyset$.

Indeed, bit strings $x$ that correspond to \emph{inconsistent} access structures are \emph{automatically} empty. The simplest example of this occurs in the holographic proof of the subadditivity of entanglement entropy:
\begin{equation}
S(A) + S(B) \geq S(AB)
\end{equation}
There, the bit string $x=11$ stands for $E(A) \cap E(B)$, which is automatically empty because entanglement wedges of disjoint regions are disjoint; viz. (\ref{ewn2}). In general, inconsistent access structures are characterized by:
\begin{equation}
\begin{array}{lp{5mm}l}
X_i \subset X_j && {\rm and}~~~x_{i} = 1~~~{\rm and}~~~x_j=0 \\
X_i \cap X_j = \emptyset &&  {\rm and}~~~x_{i} = 1~~~{\rm and}~~~x_j = 1
\end{array}
\label{eq:xinconsistent} 
\end{equation}
When the LHS of the inequality contains pairs $X_i \subset X_j$ or $X_i \subset \overline{X_j}$, the map $f$ need not be defined on all of $\{0,1\}^L$, but only on bit strings that avoid (\ref{eq:xinconsistent}) \cite{Avis:2021xnz,lhsconflicts}. Note that this restriction automatically sets $x_i = x_j$ if $X_i = X_j$, that is when the inequality features a term with a nontrivial coefficient.

\paragraph{Range of $f$: Candidate entanglement wedges} The values of $f(x) \in \{0, 1\}^R$ determine candidate entanglement wedges for regions $Y_j$ on the RHS of inequality~(\ref{eq:schematic}). These candidate wedges are bulk domains of dependence of candidate submanifolds $E^{\rm can}(Y_j)$ of the equal time slice. The latter are assembled from the $W(x)$ in a manner that mimics equation~(\ref{ewtautology}), but with respect to the $j^{\rm th}$ bit of $f(x)$:
\begin{equation}
E^{\rm can}(Y_j) = \cup_{x|f(x)_j=1} W(x)
\label{defewcandidate}
\end{equation}
These candidate objects (the wedge and its equal time slice) are generally distinct from their physical realizations. However, as we explain in Appendix~\ref{sec:oldresult}, they \emph{are} the same when inequality~(\ref{eq:schematic}) is saturated. This fact lies at the root of our results.

Object~(\ref{defewcandidate}) is a candidate entanglement wedge only if the bulk part of its boundary is homologous to $Y_j$. A simpler way of saying this is that (\ref{defewcandidate}) touches the asymptotic boundary precisely on $Y_j$ and nowhere else. To satisfy this requirement, one imposes boundary conditions on $f$. The boundary conditions concern those bit strings $x$, whose corresponding $W(x)$ touch the asymptotic boundary.

\paragraph{Boundary conditions}
The bit strings $x$ whose $W(x)$ touch the asymptotic boundary can be labeled by the atomic (indivisible) regions. For example, in the case of the monogamy of mutual information~(\ref{eq:mmi}), the atomic regions are $A$, $B$, $C$, and the purifier $O$ ($=\overline{ABC}$ in a pure state). It is easy to see that $W(x)$ meets the asymptotic boundary on $A$ and nowhere else if and only if $x = x^A$ defined by:
\begin{equation}
(x^A)_i = \begin{cases} 1 & {\rm if}~A \subset X_i \\ 0 & {\rm otherwise}\end{cases}
\label{xindicators}
\end{equation}
This is the {\bf indicator function} $x^A$ of the atomic region $A$ with respect to the composite regions $X_i$. In general, the $W(x)$s that touch the asymptotic boundary all come from indicator functions $x^A$, $x^B$ etc. of atomic regions, including the purifier.

By the same logic that relates (\ref{ewtautology}) to (\ref{defewcandidate}), we write down the analogous indicator function $f^A$ with respect to the RHS regions $Y_j$:
\begin{equation}
(f^A)_j = \begin{cases} 1 & {\rm if}~A \subset Y_j \\ 0 & {\rm otherwise}\end{cases}
\label{fxindicators}
\end{equation}
Then the candidate entanglement wedges defined in (\ref{defewcandidate}) satisfy the homology condition if and only if $f$ satisfies the {\bf boundary conditions}:
\begin{equation}
f(x^A) = f^A
\label{defbc}
\end{equation}
and likewise for all other atomic regions $B$, $C$ etc., including the purifier $O$.

\paragraph{How contraction maps prove inequalities} The defining characteristic of the RT surface of $X_i$ is that it separates the $W(x)$ with $x_i = 1$ from the $W(x)$ with $x_i=0$. This is most easily recognized from equation~(\ref{ewtautology}). As a consequence, we may decompose the area of the RT surface of $X_i$ into segments
\begin{equation}
S(X_i) = \frac{1}{2} \sum_{x, x'} |x_i - x'_i|\, {\rm Area}(x,x')\,,
\end{equation} 
where the segment with ${\rm Area}(x,x')$ is the common border of $W(x)$ and $W(x')$. Then the LHS of inequality~(\ref{eq:schematic}) takes the simple form:
\begin{equation}
\sum_{i=1}^L S(X_i) = \frac{1}{2} \sum_{x, x'} {\rm Area}(x,x') \sum_{i=1}^L |x_i - x'_i| 
\label{lhsdecomposed}
\end{equation}
Once again, we can immediately write down an analogue of this expression, which pertains to the candidate entanglement wedges defined in (\ref{defewcandidate}). Here, instead of actual RT surfaces whose areas compute $S(Y_j)$, we are characterizing candidate RT surfaces, which form the bulk part of the boundary of $E^{\rm can}(Y_j)$:
\begin{equation}
\sum_{j=1}^R S^{\rm can} (Y_j) = \frac{1}{2} \sum_{x, x'} {\rm Area}(x,x') \sum_{i=1}^R |f(x)_j - f(x')_j| 
\label{rhscandecomposed}
\end{equation}
If map $f$ satisfies the {\bf contraction condition}
\begin{equation}
\sum_{i=1}^L |x_i - x'_i| \geq \sum_{i=1}^R |f(x)_j - f(x')_j| \qquad \textrm{for all}~x, x' \in \{0,1\}^L
\label{defcontraction}
\end{equation}
then inequality~(\ref{eq:schematic}) is proven because:
\begin{equation}
\sum_{i=1}^L S(X_i) 
~\stackrel{(\flat)}{\geq}~ \sum_{j=1}^R S^{\rm can}(Y_j) 
~\stackrel{(\sharp)}{\geq}~ \sum_{j=1}^R S(Y_j)
\label{sandwich}
\end{equation}
Step~$(\flat)$ follows by substituting~(\ref{defcontraction}) into (\ref{lhsdecomposed}-\ref{rhscandecomposed}) while step~$(\sharp)$ is manifest because the physical RT surfaces have minimal areas.

\section{A contraction-based proof of the Inclusion result}
\label{sec:anotherproof}
We give another proof of result~(\ref{eq:lemma}):
\begin{itemize}
\item Given a superbalanced inequality~(\ref{eq:schematic}) and a contraction map $f$ that proves it, for every $i$  there exists a $j$ such that $X_i \subseteq Y_j$.
\end{itemize}
In contrast to the argument in Section~\ref{sec:inclusion}, here we do not assume the conjecture that all holographic entropy inequalities pass the majorization test of Reference~\cite{Grimaldi:2025jad}. We do assume, however, that the inequality can be proved by the contraction method.

For convenience, in this proof we define:
\begin{align}
|| x - x' || & \equiv \sum_i \alpha_i |x_i - x'_i| \\
|| f(x) - f(x') || & \equiv \sum_j \beta_j |f(x)_j - f(x')_j|
\end{align}
These objects are norms on vector spaces over $\mathbb{F}_2$, in which $x$ and $f(x)$ live. They therefore obey the triangle inequality 
\begin{equation}
|| x - x'' || + ||x '' - x' || \geq || x - x' ||
\label{eq:triangle}
\end{equation}
and likewise for the norm on $f(x)$. We say that $x''$ is \emph{colinear} with $x, x'$ if (\ref{eq:triangle}) holds with equality. Colinearity is a useful concept because it imposes many constraints on contraction maps \cite{Bao:2015bfa}.

A key fact is that UV-finiteness (superbalance) implies that for any two boundary conditions (\ref{defbc}) we have \cite{He:2020xuo}:
\begin{equation}
|| x^A - x^O || = || f^A - f^O ||
\label{eq:superiso}
\end{equation}
Indeed, a Bell pair shared by $A$ and $O$---which models entangled UV degrees of freedom sitting astride their common border---would contribute $|| x^A - x^O || \log 2$ to the LHS of (\ref{eq:schematic}) and $|| f^A - f^O || \log 2$ to the RHS. In (\ref{eq:superiso}) we highlight $A$ and $O$ but the same holds for any pair of indivisible regions $B, C \ldots$. 

For any $x'$, the contraction property~(\ref{defcontraction}), the triangle inequality~(\ref{eq:triangle}) and superbalance~(\ref{eq:superiso}) imply:
\begin{align}
& || x^A - x' || + || x' - x^O || \nonumber \\
\geq\, & || f^A - f(x') || + || f(x') - f^O || \\
\geq\, & || f^A - f^O || = || x^A - x^O || \nonumber
\end{align}
When $x'$ is colinear with $x^A, x^O$, the expressions in the middle get sandwiched at $||x^A - x^O||$ and all $\geq$ signs become equalities. This leads to the following consequences (for $x'$ colinear with $x^A, x^O$):
\begin{itemize}
\item Setting the upper $\geq$ to equality:
\begin{align}
|| x^A - x' || & = ||f^A - f(x') || \nonumber \\ 
|| x' - x^O || & = || f(x') - f^O ||
\label{eq:oxp}
\end{align}
\item Setting the lower $\geq$ to equality: $f(x')$ is colinear with $f^A, f^O$.
\item The colinearity further implies that the combination $f^O_j = f^A_j = 0$ and $f(x')_j = 1$ is not possible.
\end{itemize}

It suffices to prove (\ref{eq:lemma}) for those $X_i$, which are not contained in any other $X_{i'}$. Consider any such $X_i$ and let $x' = 0\ldots 010 \ldots 0$, where the single 1 is in the $i^{\rm th}$ position. The stipulation that $X_i$ is not contained in any other $X$-region is necessary for otherwise $x'$ would be excluded from the domain of $f$ due to the caveat (\ref{eq:xinconsistent}). 

Examine any fundamental region, which is contained in $X_i$, say $A$. It has the property that $x'$ is colinear with $x^A, x^O$. Therefore $f(x')$ is colinear with $f^A$ and $f^O$. Take any $j$ such that $f(x')_j = 1$. (In commonly studied inequalities this $j$ is unique, but there can be more than one such $j$ when the coefficients $\alpha_i$ and $\beta_j$ in (\ref{eq:schematic}) are unequal.) Since $f^O_j = 0$, we conclude that $f^A_j = 1$ and therefore $A \subset Y_j$. In summary, we have found that $X_i \subseteq Y_j$ for every $j$ such that $f(x')_j = 1$. 

This proves the lemma unless $f(x') = 00 \ldots 00$. But that is impossible because of (\ref{eq:oxp}).

\section{Prior results: Inequality saturation constrains erasure correcting codes}
\label{sec:oldresult}
We review why saturating a valid entropy inequality amounts to a no-go condition on certain access structures in holographic erasure correcting codes. The original argument appeared in Reference~\cite{Czech:2025jnw}, which also contains multiple examples. As elsewhere in the paper, we assume time reversal symmetry in the bulk.

\paragraph{Candidate entanglement wedges violate nesting}
As mentioned above, the candidate entanglement wedges are generally different from the physical wedges ${\rm EW}(Y_j)$. For our purposes, of particular importance are instances where the non-equality $E^{\rm can}(Y_j) \neq E(Y_j)$ can be detected directly from the contraction map $f$.

This happens whenever map $f$ contains one of the following {\bf nesting violations}:
\begin{equation}
\begin{array}{lp{5mm}l}
X_i \supset Y_j && {\rm and}~~~x_{i} = 0~~~{\rm and}~~~f(x)_j=1 \\
X_i \subset Y_j &&  {\rm and}~~~x_{i} = 1~~~{\rm and}~~~f(x)_j = 0 \\
X_i \cap Y_j = \emptyset &&  {\rm and}~~~x_{i} = 1~~~{\rm and}~~~f(x)_j = 1
\end{array}
\label{eq:violations} 
\end{equation}
When $W(x) \neq \emptyset$, these arrangements make it impossible for $E^{\rm can}(Y_j)$ to be subsumed in the physical entanglement wedge because they are inconsistent with entanglement wedge nesting~(\ref{ewn1}-\ref{ewn2}). For example, in the middle line of (\ref{eq:violations}), $x_i = 1$ and $f(x)_j = 0$ indicate that $W(x)$ is in $E(X_i)$ but \emph{outside} $E^{\rm can}(Y_j)$ yet entanglement wedge nesting requires that all of $E(X_i)$ be \emph{contained in} $E(Y_j)$. The other two lines in (\ref{eq:violations}) disobey wedge nesting for similar reasons.

For later convenience, we highlight the contrapositive of this observation: 
\medskip
\newline
{\bf Lemma~1:} In each instance of (\ref{eq:violations}), if $E^{\rm can}(Y_j) = E(Y_j)$ then $W(x) = \emptyset$.
\medskip
\newline
That is, an erasure correcting code realizes access structures $x$ described by (\ref{eq:violations}) only if some candidate wedge slice $E^{\rm can}(Y_j)$ is distinct from the physical wedge slice $E(Y_j)$.

In Reference~\cite{Czech:2025tds}, we studied the occurrences of {nesting violations} (\ref{eq:violations}) in contraction proofs of all known holographic entropy inequalities. We found that they occur with high density in proofs of \emph{all} known holographic entropy inequalities except the monogamy of mutual information. Moreover, in cases where this can be quantified, (\ref{eq:violations}) occur with \emph{maximal density}.

\paragraph{Candidate wedges violate nesting only when inequality is not saturated} Suppose that inequality~(\ref{eq:schematic}) is saturated. Then steps $(\flat)$ and $(\sharp)$ in (\ref{sandwich}) are both equalities. This leaves out two possibilities:
\begin{itemize}
\item[(i)] All candidate entanglement wedges are physical and $E^{\rm can}(Y_j) = E(Y_j)$.
\item[(ii)] Even if this is not true for any $j$ then the bulk surfaces that demarcate $E^{\rm can}(Y_j)$ and $E(Y_j)$ both have equal and therefore minimal areas so $S^{\rm can}(Y_j) = S(Y_j)$.
\end{itemize}
Option~(ii) stipulates that the entanglement entropy of $Y_j$ is at a phase transition. In this circumstance, any region contained in $E(Y_j)$ but not in $E^{\rm can}(Y_j)$ (or vice versa) would be reconstructible from $Y_j$ (respectively: protected against erasure of $Y_j$) only marginally, rendering the code vulnerable to the tiniest perturbation of the system. Indeed, at holographic phase transitions entanglement entropies are known to be subject to enhanced corrections \cite{Dong:2020iod,Akers:2020pmf}. Thus, option~(ii) describes a scenario that is both fine tuned (phase transition) and unsuitable for constructing erasure correcting codes. For these reasons, in this paper we literally relegate option~(ii) to a footnote\footnote{In fact, a more careful treatment of option~(ii) is possible. To declare that $E^{\rm can}(Y_j) \neq E(Y_j)$, one would have to invoke some other hypothetical criterion for deciding on `the correct' entanglement wedge at a phase transition. For example, one might designate the smallest candidate (the overlap of all candidate wedges whose $S^{\rm can}(Y_j)$ achieve the minimum) to be `the correct' entanglement wedge. In an appendix of Reference~\cite{Czech:2025jnw}, we showed that consistently applying additional criteria does not change the key conclusion of the main argument being presented.} and assert simply:
\medskip
\newline
{\bf Lemma~2:} If a contraction map $f$ proves an inequality ${\rm LHS} \geq {\rm RHS}$ and the inequality is saturated (${\rm LHS} = {\rm RHS}$) then $E^{\rm can}(Y_j) = E(Y_j)$ for all RHS regions $Y_j$, modulo possible phase transitions.

\paragraph{Summary of argument:}
Suppose a bulk geometry saturates an entropy inequality, which is proven by a contraction map $f$. Then the underlying erasure correcting code cannot realize access structures $x$, which obey condition~(\ref{eq:violations}). This follows by combining Lemma~2 and Lemma~1.

\bibliographystyle{JHEP}
\bibliography{reference.bib}

@article{Maldacena:1997re,
    author = "Maldacena, Juan Martin",
    title = "{The large $N$ limit of superconformal field theories and supergravity}",
    eprint = "hep-th/9711200",
    archivePrefix = "arXiv",
    reportNumber = "HUTP-97-A097, HUTP-98-A097",
    doi = "10.4310/ATMP.1998.v2.n2.a1",
    journal = "Adv. Theor. Math. Phys.",
    volume = "2",
    pages = "231--252",
    year = "1998"
}

@article{Witten:1998qj,
    author = "Witten, Edward",
    title = "{Anti-de Sitter space and holography}",
    eprint = "hep-th/9802150",
    archivePrefix = "arXiv",
    reportNumber = "IASSNS-HEP-98-15",
    doi = "10.4310/ATMP.1998.v2.n2.a2",
    journal = "Adv. Theor. Math. Phys.",
    volume = "2",
    pages = "253--291",
    year = "1998"
}

@article{VanRaamsdonk:2010pw,
    author = "Van Raamsdonk, Mark",
    title = "{Building up spacetime with quantum entanglement}",
    eprint = "1005.3035",
    archivePrefix = "arXiv",
    primaryClass = "hep-th",
    doi = "10.1142/S0218271810018529",
    journal = "Gen. Rel. Grav.",
    volume = "42",
    pages = "2323--2329",
    year = "2010"
}

@article{Ryu:2006bv,
    author = "Ryu, Shinsei and Takayanagi, Tadashi",
    title = "{Holographic derivation of entanglement entropy from AdS/CFT}",
    eprint = "hep-th/0603001",
    archivePrefix = "arXiv",
    reportNumber = "NSF-KITP-06-11",
    doi = "10.1103/PhysRevLett.96.181602",
    journal = "Phys. Rev. Lett.",
    volume = "96",
    pages = "181602",
    year = "2006"
}

@article{rt2,
    author = "Ryu, Shinsei and Takayanagi, Tadashi",
    title = "{Aspects of holographic entanglement entropy}",
    eprint = "hep-th/0605073",
    archivePrefix = "arXiv",
    reportNumber = "NSF-KITP-06-31, KUNS-2021",
    doi = "10.1088/1126-6708/2006/08/045",
    journal = "JHEP",
    volume = "08",
    pages = "045",
    year = "2006"
}

@article{hrt,
    author = "Hubeny, Veronika E. and Rangamani, Mukund and Takayanagi, Tadashi",
    title = "{A covariant holographic entanglement entropy proposal}",
    eprint = "0705.0016",
    archivePrefix = "arXiv",
    primaryClass = "hep-th",
    reportNumber = "DCPT-07-13, KUNS-2069",
    doi = "10.1088/1126-6708/2007/07/062",
    journal = "JHEP",
    volume = "07",
    pages = "062",
    year = "2007"
}

@article{maximin,
    author = "Wall, Aron C.",
    title = "{Maximin surfaces and the strong subadditivity of the covariant holographic entanglement entropy}",
    eprint = "1211.3494",
    archivePrefix = "arXiv",
    primaryClass = "hep-th",
    doi = "10.1088/0264-9381/31/22/225007",
    journal = "Class. Quant. Grav.",
    volume = "31",
    number = "22",
    pages = "225007",
    year = "2014"
}

@article{Grado-White:2024gtx,
    author = "Grado-White, Brianna and Grimaldi, Guglielmo and Headrick, Matthew and Hubeny, Veronika E.",
    title = "{Testing holographic entropy inequalities in 2 + 1 dimensions}",
    eprint = "2407.07165",
    archivePrefix = "arXiv",
    primaryClass = "hep-th",
    reportNumber = "BRX-TH-6721",
    doi = "10.1007/JHEP01(2025)065",
    journal = "JHEP",
    volume = "01",
    pages = "065",
    year = "2025"
}

@article{Grado-White:2025jci,
    author = "Grado-White, Brianna and Grimaldi, Guglielmo and Headrick, Matthew and Hubeny, Veronika E.",
    title = "{Minimax surfaces and the holographic entropy cone}",
    eprint = "2502.09894",
    archivePrefix = "arXiv",
    primaryClass = "hep-th",
    doi = "10.1007/JHEP05(2025)104",
    journal = "JHEP",
    volume = "05",
    pages = "104",
    year = "2025"
}

@article{Engelhardt:2014gca,
    author = "Engelhardt, Netta and Wall, Aron C.",
    title = "{Quantum extremal surfaces: Holographic entanglement entropy beyond the classical regime}",
    eprint = "1408.3203",
    archivePrefix = "arXiv",
    primaryClass = "hep-th",
    doi = "10.1007/JHEP01(2015)073",
    journal = "JHEP",
    volume = "01",
    pages = "073",
    year = "2015"
}

@article{Bao:2015bfa,
    author = "Bao, Ning and Nezami, Sepehr and Ooguri, Hirosi and Stoica, Bogdan and Sully, James and Walter, Michael",
    title = "{The holographic entropy cone}",
    eprint = "1505.07839",
    archivePrefix = "arXiv",
    primaryClass = "hep-th",
    reportNumber = "CALT-TH-2015-020, IPMU15-0074, SLAC-PUB-16294, SU-ITP-15-08",
    doi = "10.1007/JHEP09(2015)130",
    journal = "JHEP",
    volume = "09",
    pages = "130",
    year = "2015"
}

@article{HernandezCuenca:2019wgh,
    author = "Hern\'andez Cuenca, Sergio",
    title = "{Holographic entropy cone for five regions}",
    eprint = "1903.09148",
    archivePrefix = "arXiv",
    primaryClass = "hep-th",
    doi = "10.1103/PhysRevD.100.026004",
    journal = "Phys. Rev. D",
    volume = "100",
    number = "2",
    pages = "026004",
    year = "2019"
}

@article{Czech:2022fzb,
    author = "Czech, Bartlomiej and Wang, Yunfei",
    title = "{A holographic inequality for $N = 7$ regions}",
    eprint = "2209.10547",
    archivePrefix = "arXiv",
    primaryClass = "hep-th",
    doi = "10.1007/JHEP01(2023)101",
    journal = "JHEP",
    volume = "01",
    pages = "101",
    year = "2023"
}

@article{Hernandez-Cuenca:2023iqh,
    author = "Hern\'andez-Cuenca, Sergio and Hubeny, Veronika E. and Jia, Hewei Frederic",
    title = "{Holographic entropy inequalities and multipartite entanglement}",
    eprint = "2309.06296",
    archivePrefix = "arXiv",
    primaryClass = "hep-th",
    reportNumber = "MIT-CTP/5610",
    doi = "10.1007/JHEP08(2024)238",
    journal = "JHEP",
    volume = "08",
    pages = "238",
    year = "2024"
}

@article{Czech:2023xed,
    author = "Czech, Bartlomiej and Shuai, Sirui and Wang, Yixu and Zhang, Daiming",
    title = "{Holographic entropy inequalities and the topology of entanglement wedge nesting}",
    eprint = "2309.15145",
    archivePrefix = "arXiv",
    primaryClass = "hep-th",
    doi = "10.1103/PhysRevD.109.L101903",
    journal = "Phys. Rev. D",
    volume = "109",
    number = "10",
    pages = "L101903",
    year = "2024"
}

@article{sergioqes,
    author = "Akers, Chris and Hern\'andez-Cuenca, Sergio and Rath, Pratik",
    title = "{Quantum extremal surfaces and the holographic entropy cone}",
    eprint = "2108.07280",
    archivePrefix = "arXiv",
    primaryClass = "hep-th",
    doi = "10.1007/JHEP11(2021)177",
    journal = "JHEP",
    volume = "11",
    pages = "177",
    year = "2021"
}

@article{He:2020xuo,
    author = "He, Temple and Hubeny, Veronika E. and Rangamani, Mukund",
    title = "{Superbalance of holographic entropy inequalities}",
    eprint = "2002.04558",
    archivePrefix = "arXiv",
    primaryClass = "hep-th",
    doi = "10.1007/JHEP07(2020)245",
    journal = "JHEP",
    volume = "07",
    pages = "245",
    year = "2020"
}

@article{Hayden:2011ag,
    author = "Hayden, Patrick and Headrick, Matthew and Maloney, Alexander",
    title = "{Holographic mutual information is monogamous}",
    eprint = "1107.2940",
    archivePrefix = "arXiv",
    primaryClass = "hep-th",
    reportNumber = "BRX-TH-638",
    doi = "10.1103/PhysRevD.87.046003",
    journal = "Phys. Rev. D",
    volume = "87",
    number = "4",
    pages = "046003",
    year = "2013"
}

@article{Almheiri:2014lwa,
    author = "Almheiri, Ahmed and Dong, Xi and Harlow, Daniel",
    title = "{Bulk locality and quantum error correction in AdS/CFT}",
    eprint = "1411.7041",
    archivePrefix = "arXiv",
    primaryClass = "hep-th",
    reportNumber = "SU-ITP-14-30",
    doi = "10.1007/JHEP04(2015)163",
    journal = "JHEP",
    volume = "04",
    pages = "163",
    year = "2015"
}

@article{Czech:2019lps,
    author = "Czech, Bartlomiej and Dong, Xi",
    title = "{Holographic entropy cone with time dependence in two dimensions}",
    eprint = "1905.03787",
    archivePrefix = "arXiv",
    primaryClass = "hep-th",
    doi = "10.1007/JHEP10(2019)177",
    journal = "JHEP",
    volume = "10",
    pages = "177",
    year = "2019"
}

@article{Czech:2012bh,
    author = "Czech, Bartlomiej and Karczmarek, Joanna L. and Nogueira, Fernando and Van Raamsdonk, Mark",
    title = "{The gravity dual of a density matrix}",
    eprint = "1204.1330",
    archivePrefix = "arXiv",
    primaryClass = "hep-th",
    doi = "10.1088/0264-9381/29/15/155009",
    journal = "Class. Quant. Grav.",
    volume = "29",
    pages = "155009",
    year = "2012"
}

@article{lhsconflicts,
    author = "Li, Nan and Dong, Chuan-Shi and Du, Dong-Hui and Shu, Fu-Wen",
    title = "{Improved proof-by-contraction method and relative homologous entropy inequalities}",
    eprint = "2204.03192",
    archivePrefix = "arXiv",
    primaryClass = "hep-th",
    doi = "10.1007/JHEP06(2022)153",
    journal = "JHEP",
    volume = "06",
    pages = "153",
    year = "2022"
}

@article{Dong:2016eik,
    author = "Dong, Xi and Harlow, Daniel and Wall, Aron C.",
    title = "{Reconstruction of bulk operators within the entanglement wedge in gauge-gravity duality}",
    eprint = "1601.05416",
    archivePrefix = "arXiv",
    primaryClass = "hep-th",
    reportNumber = "NSF-KITP-16-005",
    doi = "10.1103/PhysRevLett.117.021601",
    journal = "Phys. Rev. Lett.",
    volume = "117",
    number = "2",
    pages = "021601",
    year = "2016"
}

@article{May:2021raz,
    author = "May, Alex",
    title = "{Bulk private curves require large conditional mutual information}",
    eprint = "2105.08094",
    archivePrefix = "arXiv",
    primaryClass = "hep-th",
    doi = "10.1007/JHEP09(2021)042",
    journal = "JHEP",
    volume = "09",
    pages = "042",
    year = "2021"
}

@article{Czech:2025jnw,
    author = "Czech, Bartlomiej and Shuai, Sirui and Wang, Yixu",
    title = "{Entropy inequalities constrain holographic erasure correction}",
    eprint = "2502.12246",
    archivePrefix = "arXiv",
    primaryClass = "hep-th",
    doi = "10.1103/dl3c-h3hg",
    journal = "Phys. Rev. Lett.",
    volume = "135",
    number = "14",
    pages = "141603",
    year = "2025"
}

@article{Grimaldi:2025jad,
    author = "Grimaldi, Guglielmo and Headrick, Matthew and Hubeny, Veronika E.",
    title = "{A new characterization of the holographic entropy cone}",
    eprint = "2508.21823",
    archivePrefix = "arXiv",
    primaryClass = "hep-th",
    month = "8",
    year = "2025"
}

@article{Avis:2021xnz,
    author = "Avis, David and Hern{\'a}ndez-Cuenca, Sergio",
    title = "{On the foundations and extremal structure of the holographic entropy cone}",
    eprint = "2102.07535",
    archivePrefix = "arXiv",
    primaryClass = "math.CO",
    doi = "10.1016/j.dam.2022.11.016",
    journal = "Discrete Appl. Math.",
    volume = "328",
    pages = "16--39",
    year = "2023"
}

@article{Czech:2021rxe,
    author = "Czech, Bartlomiej and Shuai, Sirui",
    title = "{Holographic cone of average entropies}",
    eprint = "2112.00763",
    archivePrefix = "arXiv",
    primaryClass = "hep-th",
    doi = "10.1038/s42005-022-01019-6",
    journal = "Commun. Phys.",
    volume = "5",
    pages = "244",
    year = "2022"
}

@article{Czech:2025tds,
    author = "Czech, Bartlomiej and Shuai, Sirui",
    title = "{Nesting is not contracting}",
    eprint = "2501.17222",
    archivePrefix = "arXiv",
    primaryClass = "hep-th",
    doi = "10.1007/JHEP06(2025)122",
    journal = "JHEP",
    volume = "06",
    pages = "122",
    year = "2025"
}

@article{Bao:2025sjn,
    author = "Bao, Ning and Furuya, Keiichiro and Naskar, Joydeep",
    title = "{On the completeness of contraction map proof method for holographic entropy inequalities}",
    eprint = "2506.18086",
    archivePrefix = "arXiv",
    primaryClass = "hep-th",
    month = "6",
    year = "2025"
}

@article{Dong:2020iod,
    author = "Dong, Xi and Wang, Huajia",
    title = "{Enhanced corrections near holographic entanglement transitions: a chaotic case study}",
    eprint = "2006.10051",
    archivePrefix = "arXiv",
    primaryClass = "hep-th",
    doi = "10.1007/JHEP11(2020)007",
    journal = "JHEP",
    volume = "11",
    pages = "007",
    year = "2020"
}

@article{Akers:2020pmf,
    author = "Akers, Chris and Penington, Geoff",
    title = "{Leading order corrections to the quantum extremal surface prescription}",
    eprint = "2008.03319",
    archivePrefix = "arXiv",
    primaryClass = "hep-th",
    doi = "10.1007/JHEP04(2021)062",
    journal = "JHEP",
    volume = "04",
    pages = "062",
    year = "2021"
}

@article{Casini:2004bw,
    author = "Casini, H. and Huerta, M.",
    title = "{A Finite entanglement entropy and the c-theorem}",
    eprint = "hep-th/0405111",
    archivePrefix = "arXiv",
    doi = "10.1016/j.physletb.2004.08.072",
    journal = "Phys. Lett. B",
    volume = "600",
    pages = "142--150",
    year = "2004"
}

@article{Myers:2010xs,
    author = "Myers, Robert C. and Sinha, Aninda",
    title = "{Seeing a c-theorem with holography}",
    eprint = "1006.1263",
    archivePrefix = "arXiv",
    primaryClass = "hep-th",
    doi = "10.1103/PhysRevD.82.046006",
    journal = "Phys. Rev. D",
    volume = "82",
    pages = "046006",
    year = "2010"
}

@article{Casini:2012ei,
    author = "Casini, H. and Huerta, Marina",
    title = "{On the RG running of the entanglement entropy of a circle}",
    eprint = "1202.5650",
    archivePrefix = "arXiv",
    primaryClass = "hep-th",
    doi = "10.1103/PhysRevD.85.125016",
    journal = "Phys. Rev. D",
    volume = "85",
    pages = "125016",
    year = "2012"
}

@article{Casini:2016udt,
    author = "Casini, Horacio and Teste, Eduardo and Torroba, Gonzalo",
    title = "{Relative entropy and the RG flow}",
    eprint = "1611.00016",
    archivePrefix = "arXiv",
    primaryClass = "hep-th",
    doi = "10.1007/JHEP03(2017)089",
    journal = "JHEP",
    volume = "03",
    pages = "089",
    year = "2017"
}

@article{Casini:2017vbe,
    author = "Casini, Horacio and Test{\'e}, Eduardo and Torroba, Gonzalo",
    title = "{Markov Property of the Conformal Field Theory Vacuum and the a Theorem}",
    eprint = "1704.01870",
    archivePrefix = "arXiv",
    primaryClass = "hep-th",
    doi = "10.1103/PhysRevLett.118.261602",
    journal = "Phys. Rev. Lett.",
    volume = "118",
    number = "26",
    pages = "261602",
    year = "2017"
}

@article{Balasubramanian:1999jd,
    author = "Balasubramanian, Vijay and Kraus, Per",
    title = "{Space-time and the holographic renormalization group}",
    eprint = "hep-th/9903190",
    archivePrefix = "arXiv",
    reportNumber = "HUTP-99-A016, EFI-99-9, NSF-ITP-99-18",
    doi = "10.1103/PhysRevLett.83.3605",
    journal = "Phys. Rev. Lett.",
    volume = "83",
    pages = "3605--3608",
    year = "1999"
}

@article{Vidal:2006sxo,
    author = "Vidal, Guifre",
    title = "{Entanglement Renormalization}",
    eprint = "cond-mat/0512165",
    archivePrefix = "arXiv",
    doi = "10.1103/PhysRevLett.99.220405",
    journal = "Phys. Rev. Lett.",
    volume = "99",
    number = "22",
    pages = "220405",
    year = "2007"
}

@article{Vidal:2008zz,
    author = "Vidal, G.",
    title = "{Class of Quantum Many-Body States That Can Be Efficiently Simulated}",
    eprint = "quant-ph/0610099",
    archivePrefix = "arXiv",
    doi = "10.1103/PhysRevLett.101.110501",
    journal = "Phys. Rev. Lett.",
    volume = "101",
    pages = "110501",
    year = "2008"
}

@article{Balasubramanian:2011wt,
    author = "Balasubramanian, Vijay and McDermott, Michael B. and Van Raamsdonk, Mark",
    title = "{Momentum-space entanglement and renormalization in quantum field theory}",
    eprint = "1108.3568",
    archivePrefix = "arXiv",
    primaryClass = "hep-th",
    reportNumber = "UPR-1233-T, UPR-1233-T",
    doi = "10.1103/PhysRevD.86.045014",
    journal = "Phys. Rev. D",
    volume = "86",
    pages = "045014",
    year = "2012"
}

@article{Banados:1992wn,
    author = "Banados, Maximo and Teitelboim, Claudio and Zanelli, Jorge",
    title = "{The Black hole in three-dimensional space-time}",
    eprint = "hep-th/9204099",
    archivePrefix = "arXiv",
    reportNumber = "PRINT-92-0151 (CHILE), IASSNS-HEP-92-29",
    doi = "10.1103/PhysRevLett.69.1849",
    journal = "Phys. Rev. Lett.",
    volume = "69",
    pages = "1849--1851",
    year = "1992"
}

@article{Gao:2016bin,
    author = "Gao, Ping and Jafferis, Daniel Louis and Wall, Aron C.",
    title = "{Traversable Wormholes via a Double Trace Deformation}",
    eprint = "1608.05687",
    archivePrefix = "arXiv",
    primaryClass = "hep-th",
    doi = "10.1007/JHEP12(2017)151",
    journal = "JHEP",
    volume = "12",
    pages = "151",
    year = "2017"
}

@article{Czech:2024rco,
    author = "Czech, Bartlomiej and Liu, Yu and Yu, Bo",
    title = "{Two infinite families of facets of the holographic entropy cone}",
    eprint = "2401.13029",
    archivePrefix = "arXiv",
    primaryClass = "hep-th",
    doi = "10.21468/SciPostPhys.17.3.084",
    journal = "SciPost Phys.",
    volume = "17",
    pages = "084",
    year = "2024"
}

@article{Casini:2019kex,
    author = "Casini, Horacio and Huerta, Marina and Mag{\'a}n, Javier M. and Pontello, Diego",
    title = "{Entanglement entropy and superselection sectors. Part I. Global symmetries}",
    eprint = "1905.10487",
    archivePrefix = "arXiv",
    primaryClass = "hep-th",
    doi = "10.1007/JHEP02(2020)014",
    journal = "JHEP",
    volume = "02",
    pages = "014",
    year = "2020"
}

\end{document}